\documentclass[twocolumn,showpacs]{revtex4}

\usepackage[latin1]{inputenc}
\usepackage{graphicx}%
\usepackage{dcolumn}
\usepackage{amsmath}

\makeatletter
\def\btt#1{\texttt{\@backslashchar#1}}%
\DeclareRobustCommand\bblash{\btt{\@backslashchar}}%
\makeatother

\begin{document}

 \title{Empirical  Features of Spontaneous and Induced Traffic Breakdowns in Free Flow at Highway Bottlenecks}



\author{Boris S. Kerner$^1$,     Micha Koller$^2$,      
 Sergey L. Klenov$^3$, Hubert Rehborn$^2$, and Michael Leibel$^4$}

 \affiliation{$^1$ Physik von Transport und Verkehr, Universit{\"a}t Duisburg-Essen,
47048 Duisburg, Germany}

 \affiliation{$^2$ Daimler AG,  71063 Sindelfingen, Germany }

\affiliation{$^3$ Moscow Institute of Physics and Technology, Department of Physics, 141700 Dolgoprudny,
Moscow Region, Russia}   

 \affiliation{$^4$ Karlsruhe University of Applied Sciences, 76133 Karlsruhe, Germany}



\pacs{89.40.-a, 47.54.-r, 64.60.Cn, 05.65.+b}

\begin{abstract} 
Based on an empirical study of real field  
traffic data measured  in 1996--2014 through road detectors installed on German freeways, 
we reveal physical features of empirical nuclei for spontaneous traffic breakdown in free flow
at  highway bottlenecks. It is shown that the source of a nucleus  for traffic breakdown is   the solely difference
 between
 empirical spontaneous and induced traffic breakdowns at a highway bottleneck.
    Microscopic traffic simulations with a stochastic traffic flow model in the framework of three-phase theory
  explain the empirical findings.
It turns out that in the most cases, a nucleus for empirical spontaneous traffic breakdown occurs
through an interaction of one of  waves in free flow with an empirical permanent   speed disturbance localized at 
  a highway bottleneck. 
The wave is a localized structure in free flow,
 in which the total flow rate is larger and
 the   speed averaged across the highway is smaller    than outside the wave.
  The waves in free flow   appear
  due to oscilations in the percentage of slow   vehicles; 
these  waves  
  propagate with the average speed of   slow   vehicles in free flow (about 85--88 km/h for German highways).
Any of the waves  exhibits
 a  two-dimensional   asymmetric spatiotemporal  structure:  Wave's characteristics 
 are different in different highway lanes.
  \end{abstract}

\maketitle

\section{Introduction}

In many equilibrium (e.g.,~\cite{Pound,Sanz}) and dissipative metastable systems of natural science
(e.g.,~\cite{Haken1977,Thom1975,Gardiner,NP,Scholl1987,Vasilev,Michailov1,Michailov2,KO,Chandrasekhar1961,NIE95})
there can be a spontaneous phase transition from one metastable phase to
another metastable phase of a system. 
Such   spontaneous phase transition occurs when a nucleus for the transition 
appears randomly   in an initial metastable phase of the system:
The growth of the nucleus leads to the phase transition.
The nucleus can be a fluctuation within the initial system phase
whose amplitude is equal or larger than an amplitude of a   critical nucleus 
required for   spontaneous phase transition. 
Nuclei for such spontaneous phase transitions can be observed in empirical and experimental studies of many
equilibrium and dissipative metastable systems (e.g.,~\cite{Pound,Sanz,Vasilev,Michailov1,Michailov2,KO,Chandrasekhar1961,NIE95}).
There can  also  be another source
for the occurrence of a nucleus, rather than   fluctuations:
A nucleus can  be induced by an external disturbance applied 
to the initial  phase. In this case,
the phase transition is called an induced phase transition  (e.g.,~\cite{Thom1975,NP,Vasilev,Michailov1,Michailov2,KO,Chandrasekhar1961,NIE95}).

 \begin{figure}
\begin{center}
\includegraphics*[width=10 cm]{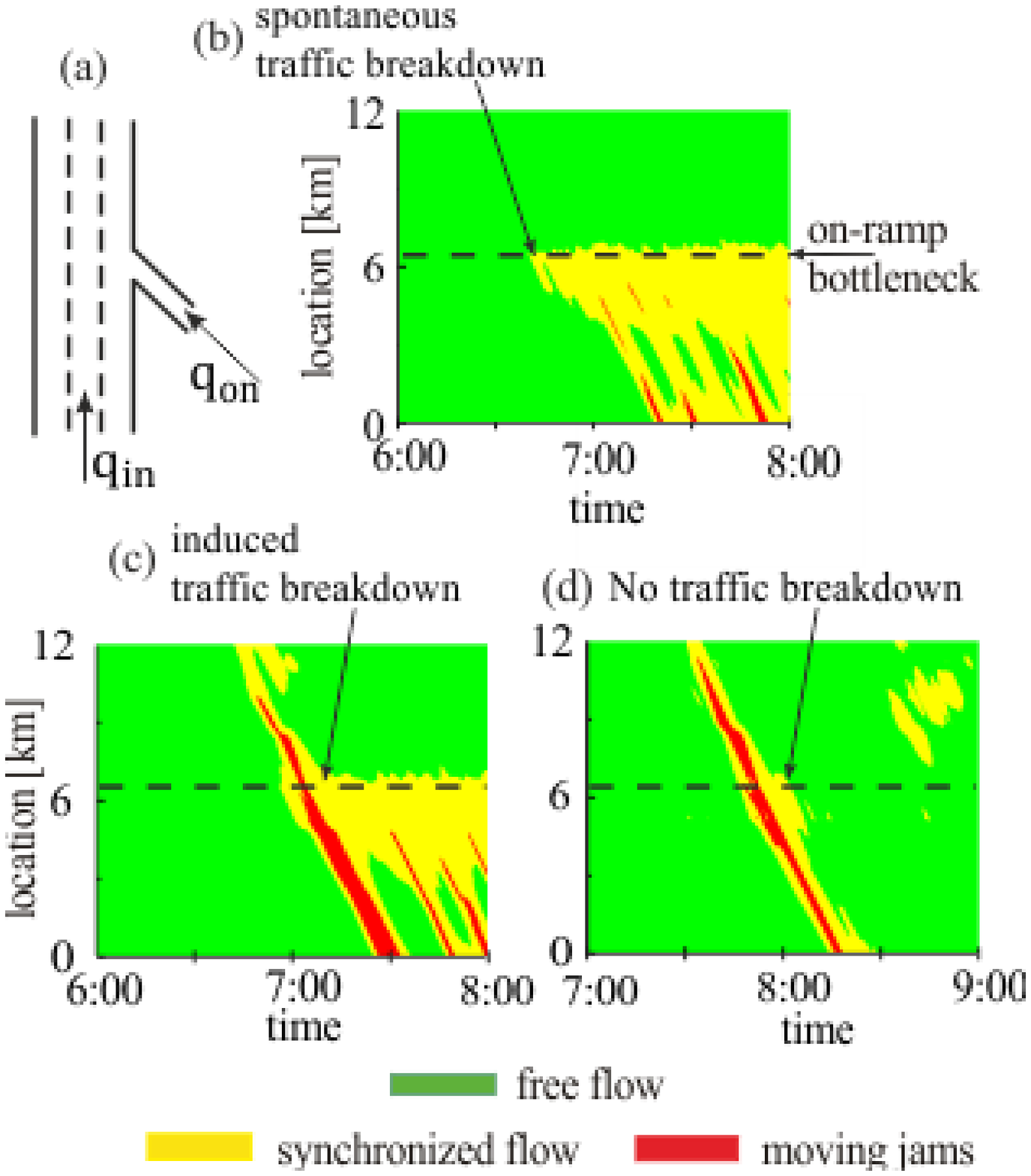}
\end{center}
\caption[]{Overview of empirical features of traffic breakdown (F$\rightarrow$S transition) at an on-ramp bottleneck:
(a) Sketch of 
   section of three-lane highway in Germany with an on-ramp bottleneck. (b--d)
Speed  data  measured with road detectors installed 
 along    road section in (a);   data is
presented in space and time with   averaging   method described in Sec.~C.2 of~\cite{KernerRSch2013}.
(b) Empirical spontaneous traffic breakdown. (c) Empirical induced traffic breakdown.
(d) Moving jam propagation through the bottleneck without induced traffic breakdown.
Real field traffic data measured by road detectors
on three-lane freeway A5-South in Germany on April 15, 1996 (b), March 22, 2001 (c), and June 23, 1998 (d).
On-ramp bottleneck  marked by dashed lines in (b--d) is   effective on-ramp bottleneck $B_{3}$ explained in detailed in Sec.~9.2.1
of~\cite{KernerBook}. Road detectors, at which 1-min  traffic data averaged across the road
  presented in (b--d) have been measured,  
are   at  locations: $x=$ 0, 1.7, 3.2, 4.7, 5.1, 6.4, 7.9, 8.8, 9.8, 11.1, 12.2 km.
}
\label{Breakdown}
\end{figure}

The occurrence of   congestion in vehicular traffic results 
  either from empirical spontaneous   or   induced traffic breakdown  
at a highway bottleneck (Fig.~\ref{Breakdown})~\cite{KernerBook,KernerBook2,Kerner_Review}.
An empirical spontaneous traffic breakdown occurs when free flow has been both upstream and downstream before
the breakdown has occurred  (Fig.~\ref{Breakdown} (b)). Empirical induced traffic breakdown
is caused by a propagation of a localized congested pattern to the bottleneck location; in the case shown in Fig.~\ref{Breakdown} (c), this localized congested pattern is a moving jam:
After the jam is far away upstream of the bottleneck, congested traffic remains at the bottleneck for a long time interval.
The downstream front
of congested traffic, which
 separates free flow downstream and congested traffic upstream of the bottleneck, is fixed
at the bottleneck.
Congested traffic whose downstream front is fixed at the bottleneck is called synchronized flow:
Traffic breakdown is a transition from free flow to synchronized flow
at the bottleneck (F$\rightarrow$S transition)~\cite{KernerBook,KernerBook2}.

Because  there can  be either empirical spontaneous or induced traffic breakdown at the bottleneck  
(Fig.~\ref{Breakdown} (b, c)), in three-phase traffic theory is assumed that
under conditions $C_{\rm min}\leq q_{\rm sum} < C_{\rm max}$  free flow is in a metastable state with respect to
an F$\rightarrow$S transition at the bottleneck, where $q_{\rm sum}$ is the flow rate in free flow at the bottleneck, $C_{\rm min}$ and $C_{\rm max}$ are, respectively, some
  minimum capacity and maximum capacity
 of free flow at the bottleneck~\cite{KernerBook,KernerBook2,Kerner_Review}.  
 This means that in an empirical example shown
in  Fig.~\ref{Breakdown} (d), in which 
 due to jam propagation through the bottleneck {\it no} traffic breakdown has  been induced
at the bottleneck, condition 
  $q_{\rm sum}<C_{\rm min}$ should be satisfied.
 
As in other metastable systems, we could expect that when an empirical spontaneous traffic breakdown is observed,
there   should also be   a disturbance  in free flow   
that acts as a nucleus
for the F$\rightarrow$S transition (traffic breakdown) at the bottleneck.
However, up to now no  nuclei, which    are responsible for spontaneous traffic breakdown at highway bottlenecks,
could be identified in real field traffic data measured in free flow.
In this article, we reveal  empirical
nuclei for spontaneous traffic breakdown in free flow
at highway bottlenecks and study their physics.

The article is organized as follows.
 The physics of empirical   nuclei for
spontaneous traffic breakdown at  highway bottlenecks is
  the subject of Sec.~\ref{Nuclei_S}. An empirical spatiotemporal  structure of nuclei   and a
  microscopic theory
 of the nucleation of traffic breakdown at highway bottlenecks are considered in Sec.~\ref{2D_Theory_S}.
 In  Sec.~\ref{Dis_S},  we discuss
   empirical features of different sources of nuclei for empirical   traffic breakdown at highway bottlenecks (Secs.~\ref{Ind_Sp_FS_S} and~\ref{Spillover_S}) as well as
formulate conclusions.

 \section{Physics of Empirical Nucleation of Traffic Breakdown at Highway Bottlenecks \label{Nuclei_S} }

\subsection{Methodology of study of waves in empirical free flow \label{Wave_Sub_S}}

In each of the freeway lanes, road detectors measure  the following 1-min averaged data:
the flow rate of all vehicles $q$, the flow rate of long vehicles $q_{\rm slow}$, and the  
average speed $v$; respectively, we can calculate the percentage of long vehicles   $\psi=100 q_{\rm slow}/q$ that can be considered {\it slow vehicles}
because 
the most of long vehicles moving on working days on German highways have a speed limit 80 km/h (in reality,   slow vehicles move usually at the 
speed within a range 80--90 km/h).

To find nuclei for traffic breakdown, we study possible waves   in free flow.  Additionally with 
possible waves of 1-min average traffic variables $q$, $q_{\rm slow}$,  $v$, and $\psi$,
 we inverstigate also waves of the following variables
\begin{eqnarray}
\label{wave_variable}
\Delta q_{\rm wave}= q - \bar q, \
 \Delta \psi_{\rm wave}= \psi - \bar \psi, \ {\rm and} \\ 
  \Delta v_{\rm wave}= \bar v -  v, \nonumber
\end{eqnarray}
where     
traffic variables $\bar q, \ \bar v, \ \bar \psi$ are related  to 20-min average data with the used of the well-known procedure of 
$\lq\lq$moving averaging".

\begin{figure}
\begin{center}
\includegraphics*[width=10 cm]{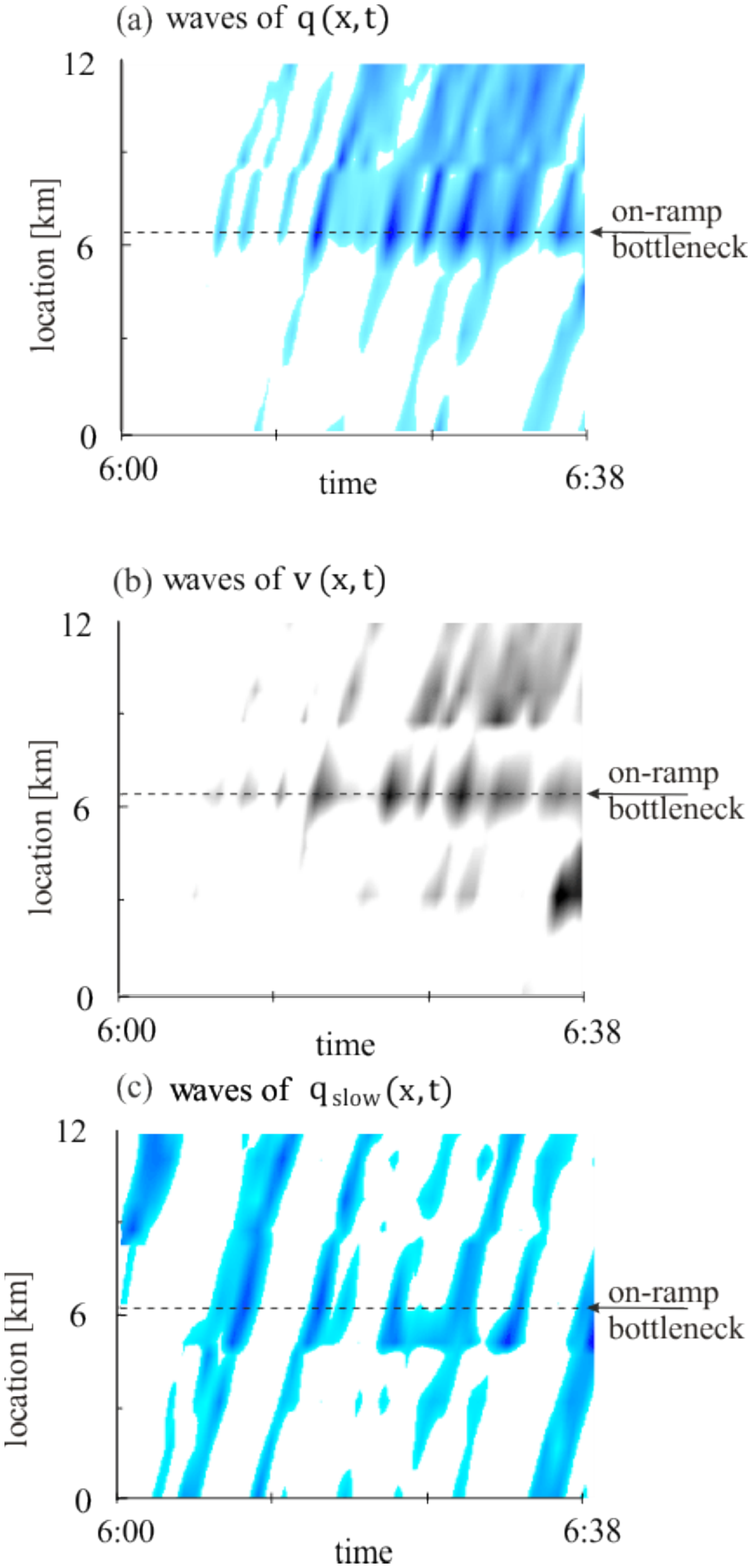}
\end{center}
\caption[]{Empirical waves in free flow    
shown for time interval $06:00  \leq  t  \leq 06:38$
 before traffic breakdown has occurred at $ t = 06:39$.
  Real field traffic data measured by road detectors   on April 15, 1996 (Monday) (Fig.~\ref{Breakdown} (b)):    (a)   
Waves of the total flow rate   $q(x, t)$ 
 presented  by regions with 
variable shades of gray (blue   in the on-line version)  (in white regions $q \leq$ 5400 vehicles/h,
in black (dark blue) regions $q\geq$ 8000 vehicles/h). (b)
Waves of the  speed $v(x, t)$ averaged across the road presented  by regions with 
variable shades of gray    (in white regions $v \geq$ 100  km/h,
in black  regions $v\leq$ 75 km/h). (c) Waves of the total flow rate of long vehicles $q_{\rm slow}(x, t)$ 
 presented  by regions with 
variable shades of gray (blue   in the on-line version)  (in white regions $q_{\rm slow} \leq$ 720 vehicles/h,
in black (dark blue) regions $q_{\rm slow} \geq$ 1500 vehicles/h).
In formula (\ref{var_traffic_w}), we use $v_{\rm d}=$ 90 km/h,
number of virtual road locations $K=$ 65 between each pair of detectors,
number of virtual time steps within 1 min time interval    between two consequent measurements at road detectors is
equal to 14.
}
\label{FreeWaves1996}
\end{figure}

\begin{figure}
\begin{center}
\includegraphics*[width=10 cm]{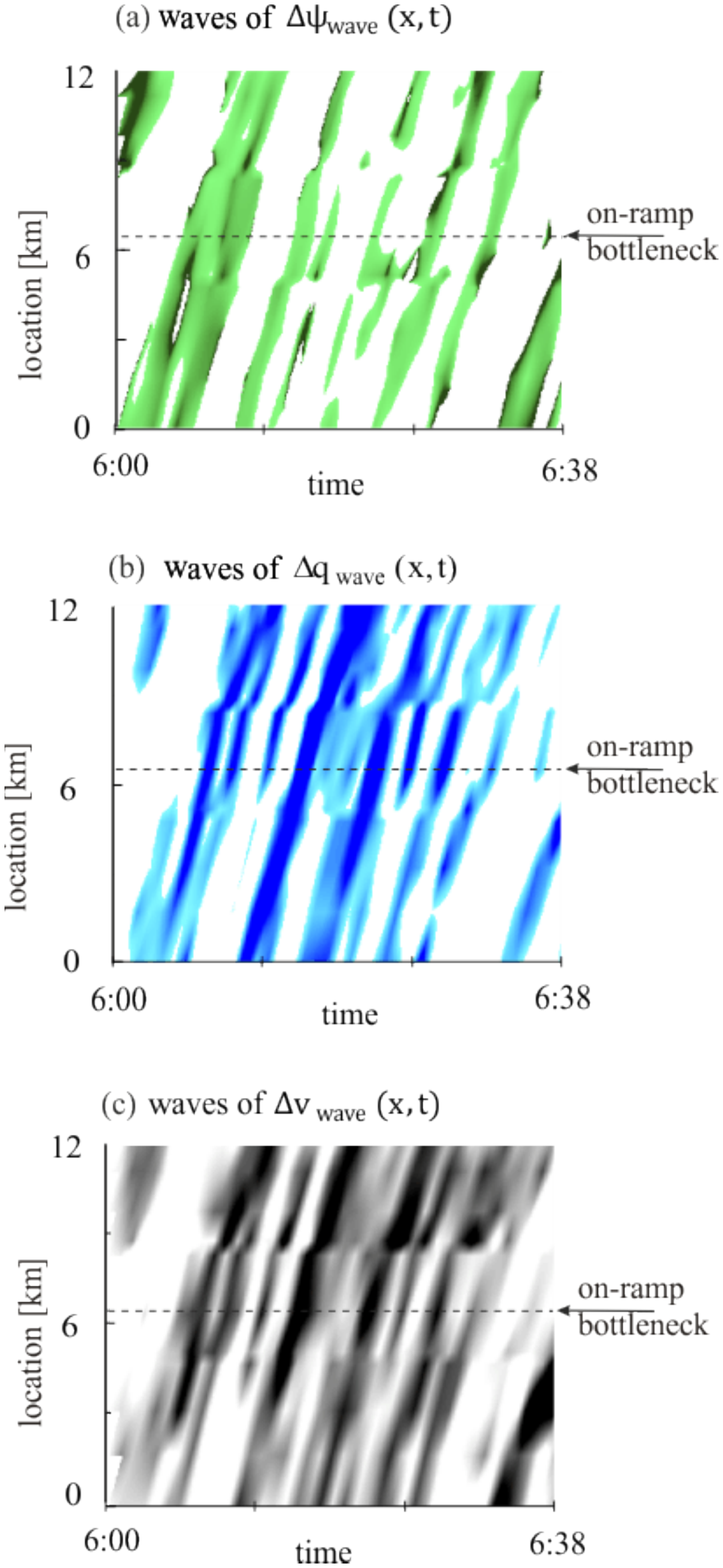}
\end{center}
\caption[]{Empirical waves of   $\Delta \psi_{\rm wave}$ (a),
  $\Delta   q_{\rm wave}$ (b), and   $\Delta   v_{\rm wave}$ (c)
 for   data    in Fig.~\ref{Breakdown} (b) averaged across the road
for the same time interval as that in Fig.~\ref{FreeWaves1996} 
  (real field traffic data measured by road detectors installed along three-lane freeway):    (a) Waves of
  $\Delta \psi_{\rm wave}(x, t)$ are presented  by regions with 
variable shades of gray (green   in the on-line version)  (in white regions $\Delta \psi_{\rm wave}\leq$ 0.1 $\%$,
in black (dark green) regions $\Delta \psi_{\rm wave}\geq$ 1 $\%$). (b)
Waves of   $\Delta q_{\rm wave}(x, t)$ are
 presented  by regions with 
variable shades of gray (blue   in the on-line version)  (in white regions $\Delta q_{\rm wave}\leq$ 600 vehicles/h,
in black (dark blue) regions $\Delta q_{\rm wave}\geq$ 1500 vehicles/h). (c)
Waves of   $\Delta v_{\rm wave}(x, t)$  are presented  by regions with 
variable shades of gray    (in white regions $\Delta v_{\rm wave}\leq$ 1 km/h,
in black  regions $\Delta v_{\rm wave} \geq$ 15  km/h).
Model parameters in formula (\ref{var_traffic_w}) are the same as those in Fig.~\ref{FreeWaves1996}.}
\label{Nuclei1996}
\end{figure}

Furthermore, to reconstruct a possible wave propagation in space and time,
we consider a pair of road detectors
whose co-ordinates
  are   $x=x_{\rm up}$   (upstream detector) and  $x_{\rm down}$ (downstream). We denote  
traffic variables measured by these detectors   by $\varphi(x_{\rm up}, \ t)$  and $\varphi(x_{\rm down}, \ t)$
for the upstream and downstream detectors, respectively, i.e., $\varphi$ denotes one of traffic variables:
\begin{equation}
\varphi  = [\Delta q_{\rm wave}, \ \Delta v_{\rm wave}, \ \Delta \psi_{\rm wave}, \ q, \ q_{\rm slow}, \ v,  \  \psi].
\label{Var_for}
\end{equation} 
Within road locations $x$ between these two detectors   
\begin{equation}
 x_{\rm up}<x<x_{\rm down},
 \label{range_co}
\end{equation}
we introduce $K$ {\it virtual} road locations ($K\gg 1$) with co-ordinates $x_{i}$ that are at a small distance $\Delta d$ each from
another, where
 $i=1,2, ..., K$,
    $\Delta d=(x_{\rm down}-x_{\rm up})/K$.  
 Then
 traffic variables $\varphi$ (\ref{Var_for})
  at each of   locations   $x_{i}$  are found from formula:
\begin{eqnarray}
\varphi(x_{i}, t)=\frac{x_{\rm down}-x_{i}}{x_{\rm down}-x_{\rm up}}\varphi(x_{\rm up}, \ t+\frac{x_{\rm up}-x_{i}}{v_{\rm d}}) +  \nonumber  \\
\frac{x_{i}-x_{\rm up}}{x_{\rm down}-x_{\rm up}}\varphi(x_{\rm down}, \ t+\frac{x_{\rm down}-x_{i}}{v_{\rm d}}),
 \label{var_traffic_w}
\end{eqnarray}
where $v_{\rm d}$ is   constant model parameter.
Some results of such  analysis of empirical waves in free flow are presented in Figs.~\ref{FreeWaves1996} and~\ref{Nuclei1996}. We have found that empirical
waves can propagate through the whole road section; some of the waves appear at on-ramps or disappear at off-ramps.

In Fig.~\ref{FreeWaves1996}, we observe a strong increase in the flow rates $q$ and $q_{\rm Lslow}$ in the flow direction that begins about 0.5--1 km 
upstream of the effective location of the on-ramp bottleneck  
 labeled by $\lq\lq$on-ramp bottleneck" in Fig.~\ref{Breakdown} (b)~\cite{FlowIncrease}.
Due to this  increase in the flow rate, the average speed decreases appreciably.

To avoid this negative impact of average values of traffic variables on wave resolution in free flow 
(Fig.~\ref{FreeWaves1996}),  in Fig.~\ref{Nuclei1996} we present  
waves of variables $\Delta \psi_{\rm wave}$,
$\Delta q_{\rm wave}$, and $\Delta v_{\rm wave}$ (\ref{wave_variable}).
Then we find out that the waves of the flow rate $\Delta q_{\rm wave}$ and
the speed $\Delta v_{\rm wave}$  
 almost  coincide with the waves of the percentage of (slow) long vehicles
$\Delta \psi_{\rm wave}$ (Fig.~\ref{Nuclei1996}).

We see that     each of the waves
of the traffic variables (\ref{Var_for}) propagates downstream with the   mean wave velocity $v_{\rm wave}$ that is approximately 
equal to   the mean speed of slow vehicles $v^{\rm (mean)}_{\rm slow}$ (Figs.~\ref{FreeWaves1996} 
and~\ref{Nuclei1996}):
\begin{equation}
v_{\rm wave}=v^{\rm (mean)}_{\rm slow}.
\label{wave_Vel}
\end{equation}
In all empirical data, $v^{\rm (mean)}_{\rm slow}$ is given by the average speed of long vehicles that changes
within   range 85--88 km/h. 
Within any of the waves propagating with the velocity $v_{\rm wave}$ (\ref{wave_Vel})
the percentage of long vehicles and the flow rate are larger, whereas the average speed is lower than outside the wave.

 \subsection{Empirical nucleation of traffic breakdown at on-ramp bottleneck}
 
\begin{figure}
\begin{center}
\includegraphics*[width=10 cm]{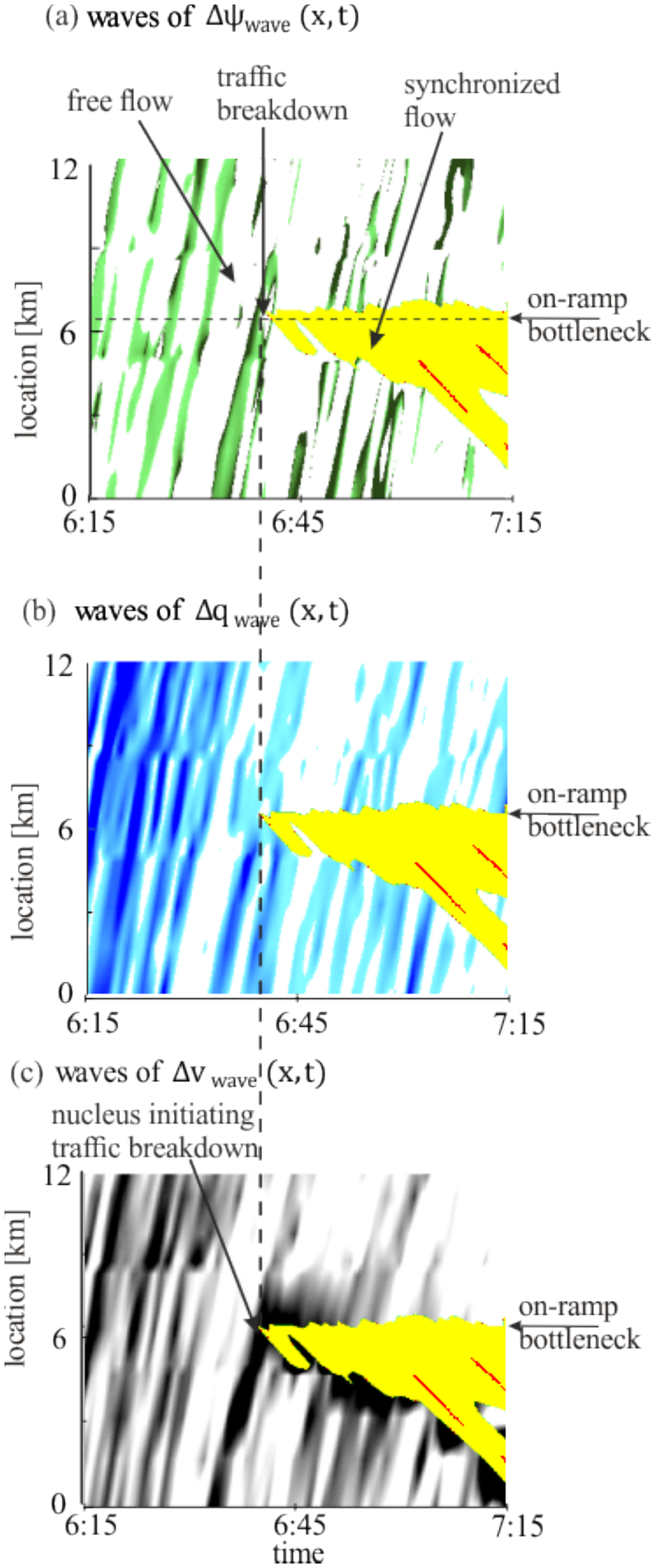}
\end{center}
\caption[]{Empirical nucleus in free flow for data     in Fig.~\ref{Breakdown} (b).
Empirical waves of   $\Delta \psi_{\rm wave}$ (a),
  $\Delta   q_{\rm wave}$ (b), and   $\Delta   v_{\rm wave}$ (c)
  in free flow for a longer time interval as that in Fig.~\ref{Nuclei1996}
  (real field traffic data measured by road detectors installed along three-lane freeway). 
In (a--c),  regions labeled by $\lq\lq$synchronized flow" show symbolically   synchronized flow.
  Parameters of the presentation of empirical waves in (a--c) are  the same as those in 
 Fig.~\ref{Nuclei1996}(a--c), respectively.}
\label{Nuclei1996_2}
\end{figure}

 During a long time interval,
 the waves of traffic variables   in free flow
 propagate with the positive
  velocity $v_{\rm wave}$ (\ref{wave_Vel})  through the on-ramp bottleneck without
  any consequences for free flow at the bottleneck (Fig.~\ref{Nuclei1996}). This   
 changes crucially when we consider a longer time interval (Fig.~\ref{Nuclei1996_2}).
 
 Indeed, when one of the waves propagates through the on-ramp bottleneck, the wave initiates
traffic breakdown at the bottleneck. During the subsequent wave propagation downstream of the bottleneck,
the structure of the wave and its features do not change.  Thus,
    one of the waves in free flow studied above becomes to be a nucleus for traffic breakdown at the bottleneck, when the wave
    propagates through the bottleneck
    (labeled by $\lq\lq$nucleus" in Fig.~\ref{Nuclei1996_2} (c)).

\subsection{Empirical nucleation of traffic breakdown at off-ramp bottleneck \label{Off-Ramp_Sec}}

\begin{figure}
\begin{center}
\includegraphics*[width=8 cm]{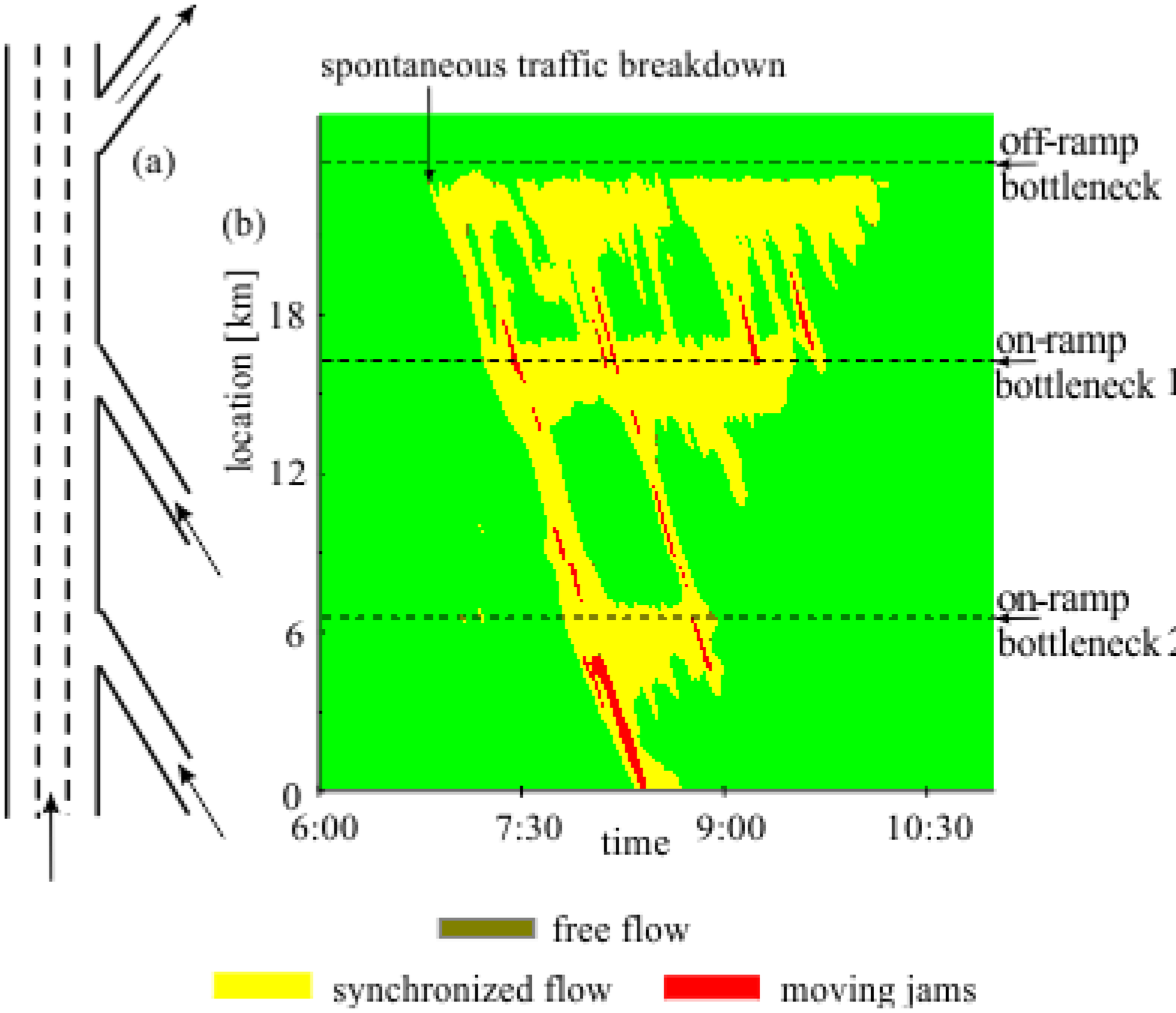}
\end{center}
\caption[]{Overview of empirical features of traffic breakdown (F$\rightarrow$S transition) at   off-ramp bottleneck:
(a) Sketch of 
   section of three-lane freeway A5-South in Germany with   off-ramp bottleneck and two upstream on-ramp bottlenecks.  (b)
Speed  data  measured with road detectors installed 
 along    road section in (a);   data is
presented in space and time with   averaging   method described in Sec.~C.2 of~\cite{KernerRSch2013}.
Real field traffic data measured by road detectors on September 03, 1998 (Thursday).
Off-ramp bottleneck, on-ramp bottleneck 1 and  on-ramp bottleneck 2 marked by dashed lines are, respectively, 
  effective   bottlenecks $B_{1}$, $B_{2}$, and $B_{3}$ explained in detailed in Sec.~9.2.1
of~\cite{KernerBook}. Road detectors   are at   locations: $x=$ 0, 1.7, 3.2, 4.7, 5.1, 6.4, 7.9, 8.8, 9.8, 11.1, 12.2, 13.7, 14.8, 15.5, 16.1, 17.0, 17.7,
18.9, 19.8, 20.8, 21.7, 22.8, 23.3, 24.0 km. The on-ramp bottleneck labeled by $\lq\lq$on-ramp bottleneck 2" in (b) is the same as that
in Fig.~\ref{Breakdown}.
}
\label{Breakdown03091998}
\end{figure}

 The
  empirical result shown  in Fig.~\ref{Nuclei1996_2} remains qualitatively the same for the case of
      wave propagation through an off-ramp bottleneck.
In   Fig.~\ref{Breakdown03091998},
there are three bottlenecks: an off-ramp bottleneck and two upstream on-ramp bottlenecks. In this case,
  traffic breakdown occurs at the off-ramp bottleneck. This traffic breakdown leads to the emergence of a complex spatiotemporal congested pattern   upstream of the off-ramp 
  bottleneck (Fig.~\ref{Breakdown03091998}).
  
\begin{figure}
\begin{center}
 \includegraphics*[width=10 cm]{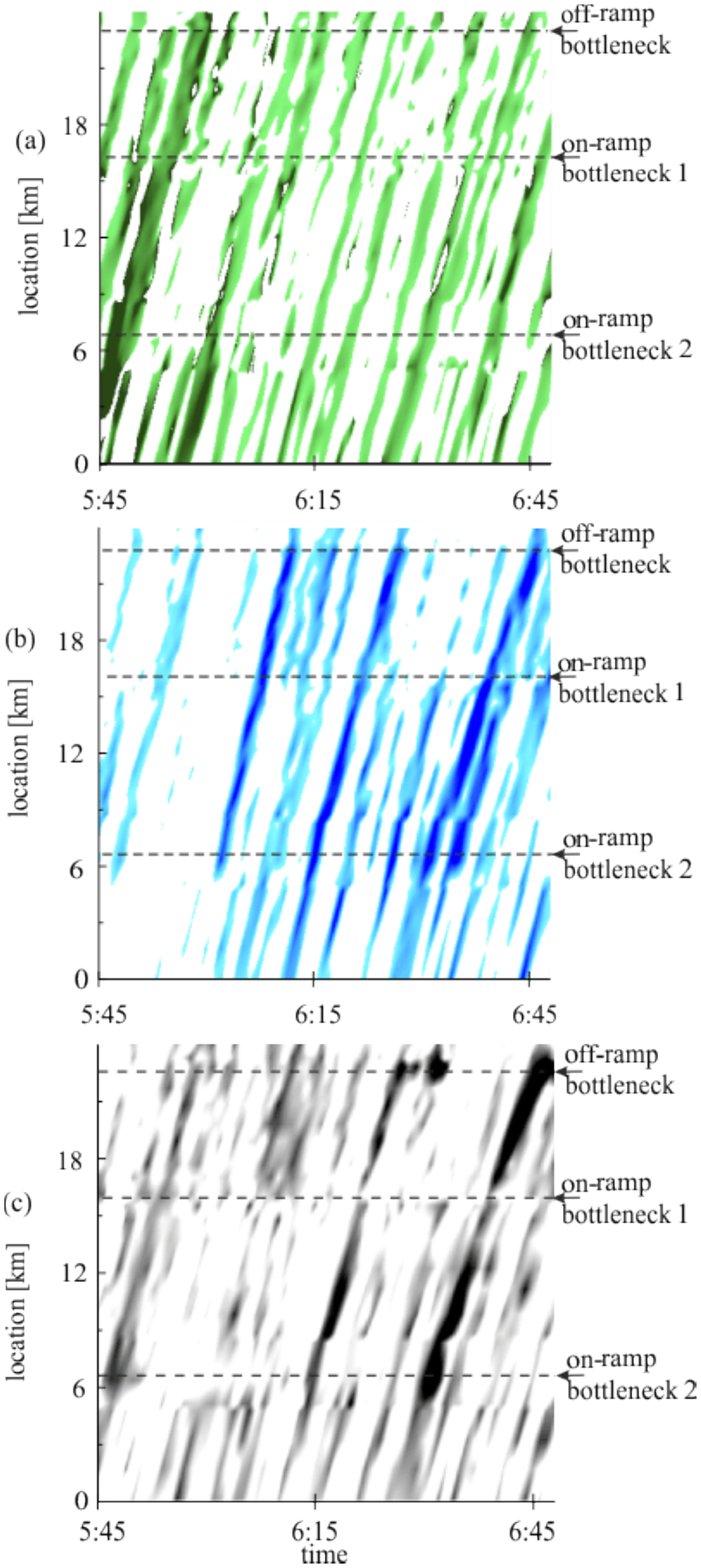}
\end{center}
\caption[]{Empirical waves in free flow averaged across the road, which are associated with  data   in Fig.~\ref{Breakdown03091998} (b),
for time interval $05:45 \leq t \leq 06:45$
 before the breakdown has  occurred:    (a) Waves of
  $\Delta \psi_{\rm wave}$ are presented  by regions with 
variable shades of gray (green   in the on-line version)  (in white regions $\Delta \psi_{\rm wave}\leq$ 0.1 $\%$,
in black (dark green) regions $\Delta \psi_{\rm wave}\geq$ 1 $\%$). (b)
Waves of   $\Delta   q_{\rm wave}$ are
 presented  by regions with 
variable shades of gray (blue   in the on-line version)  (in white regions $\Delta   q_{\rm wave}\leq$ 700 vehicles/h,
in black (dark blue) regions $\Delta   q_{\rm wave}\geq$ 2000 vehicles/h). (c)
Waves of   $\Delta   v_{\rm wave}$ are presented  by regions with 
variable shades of gray    (in white regions $\Delta   v_{\rm wave}\leq$ 2 km/h,
in black  regions $\Delta   v_{\rm wave} \geq$ 15  km/h). Real field traffic data  measured on September 03, 1998.
Model parameters in formula (\ref{var_traffic_w}) are the same as those in Fig.~\ref{FreeWaves1996}.}
\label{Nuclei1998}
\end{figure}

   Before the breakdown has occurred,   there is also a complex  sequence of waves of traffic variables $\Delta \psi_{\rm wave}$,
$\Delta q_{\rm wave}$, and $\Delta v_{\rm wave}$ in free flow; some of the waves propagate through the whole
   25 km long highway section (Fig.~\ref{Nuclei1998}).

\begin{figure}
\begin{center}
\includegraphics*[width=10 cm]{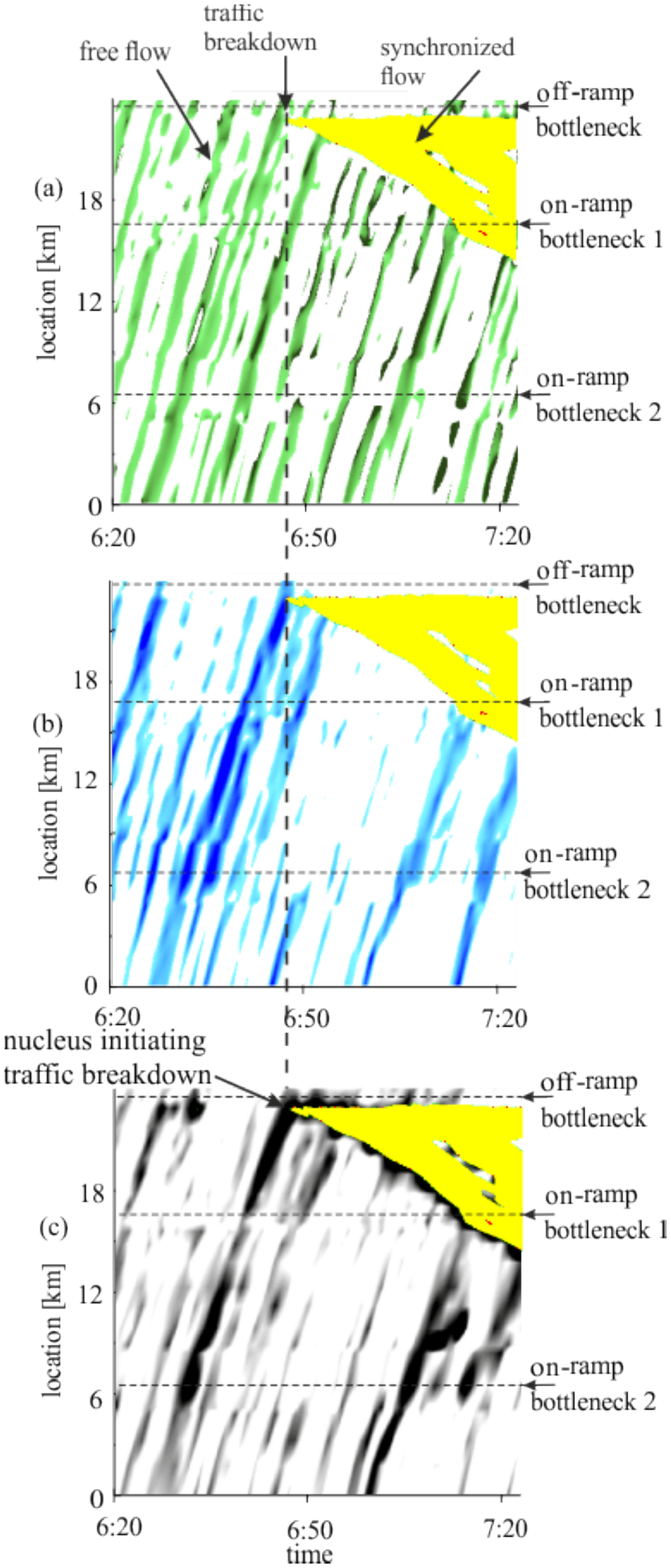}
\end{center}
\caption[]{Empirical nucleus in free flow associated with   data     in Fig.~\ref{Breakdown03091998} (b). Empirical waves of   $\Delta \psi_{\rm wave}$ (a),
  $\Delta   q_{\rm wave}$ (b), and   $\Delta   v_{\rm wave}$ (c)
  in free flow averaged across the road
for a longer time interval as that shown in Fig.~\ref{Nuclei1998}. In (a--c),  regions labeled by $\lq\lq$synchronized flow" show  symbolically   synchronized flow.
  Parameters of the presentation of empirical waves in (a--c) are  the same as those in 
 Fig.~\ref{Nuclei1998}(a--c), respectively. Real field traffic data measured by road detectors   on September 03, 1998.
}
\label{Nuclei1998_2}
\end{figure}

When we consider a longer time interval as that shown in Fig.~\ref{Nuclei1998}, we find that
 while one of the waves approaches the off-ramp bottleneck, the wave initiates
traffic breakdown at the bottleneck (Fig.~\ref{Nuclei1998_2}). The structure of the wave and its features do not change
after the wave is downstream of the off-ramp bottleneck. Thus, as in the case of the on-ramp bottleneck (Fig.~\ref{Nuclei1996_2} (c)), 
     the wave in free flow mentioned above  becomes to be a nucleus for traffic breakdown at the off-ramp bottleneck, when the wave
    propagates through this   bottleneck (Fig.~\ref{Nuclei1998_2} (c)).

\subsection{Empirical probability of spontaneous traffic breakdown at highway bottlenecks \label{Prob_S}}

In 1998, Persaud et al.~\cite{Persaud1998B} discovered that the empirical probability of traffic breakdown at   highway bottlenecks is a  
  growing flow rate function. This empirical probability of traffic breakdown has firstly been explained by an
  F$\rightarrow$S transition in a metastable
  free flow at an on-ramp bottleneck with the use of
   simulations of a three-phase cellular automaton model~\cite{KKW}. This theoretical  
     probability  of spontaneous breakdown at a highway bottleneck is well  fitted by a function~\cite{KKW}
\begin{equation}
P^{\rm (B)}=\frac{1}{1+ {\rm exp}[\alpha(q_{\rm P}-q_{\rm sum})]},
\label{Prob_For}
\end{equation}
where $q_{\rm sum}$ is the flow rate in free flow at the bottleneck, $\alpha$ and $q_{\rm P}$ are parameters. Qualitatively the same growing  flow-rate function for the breakdown probability has also been found in
measured 5-minutes average traffic data~\cite{Brilon310,Brilon210,Brilon,BrilonISTTT2009,Elefteriadou2014}.

  \begin{figure}
\begin{center}
\includegraphics*[width=10 cm]{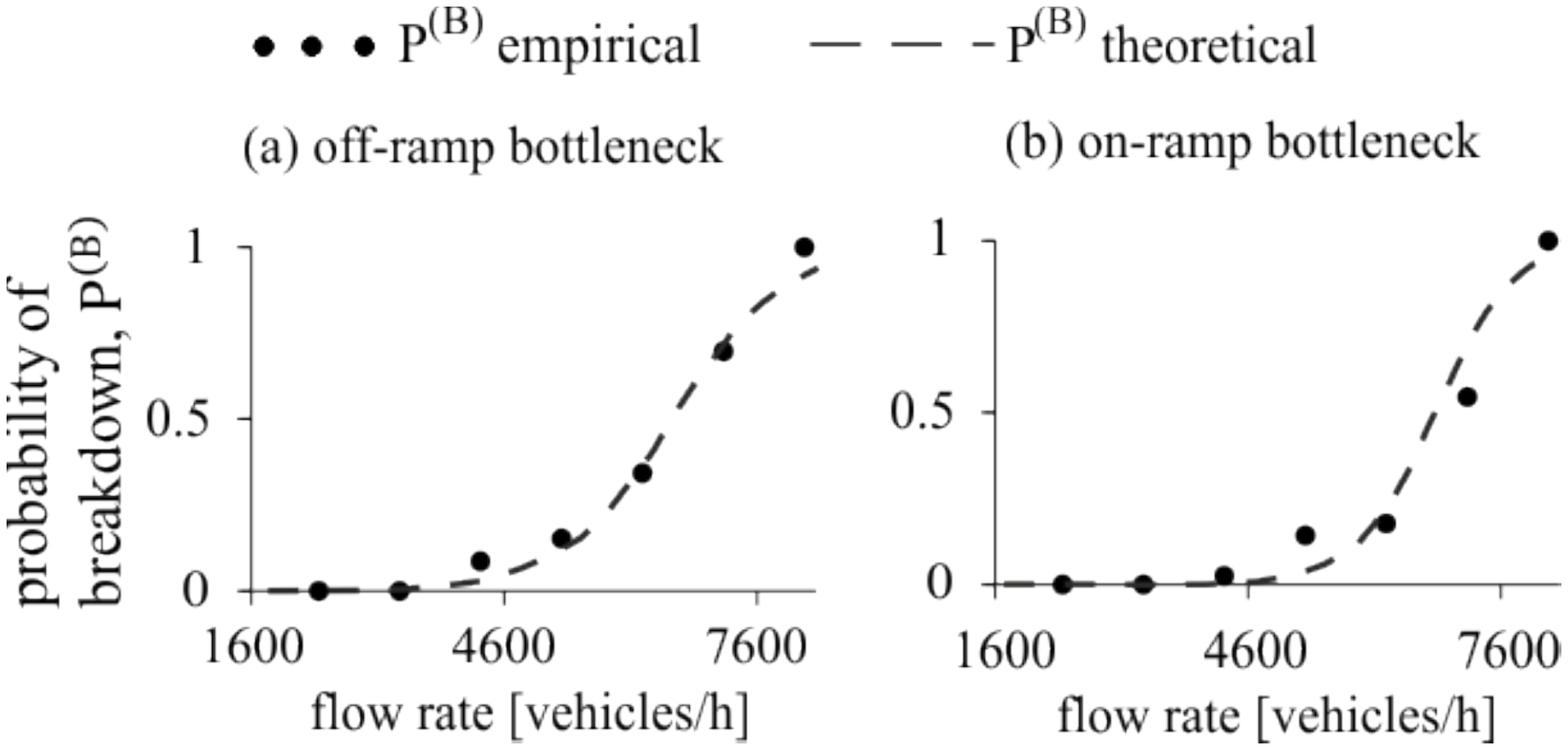}
\caption{Comparison of empirical (black points) and theoretical (dashed curves related to (\ref{Prob_For})) probabilities  
of traffic breakdown; empirical breakdown probabilities (black points) are related to real field traffic data 
measured by road detectors installed along  a section of three-lane freeway A5-South
with effective bottlenecks shown in Fig.~\ref{Breakdown03091998}: (a)
Probability of traffic breakdown at on-ramp bottleneck
(labeled by $\lq\lq$on-ramp bottleneck 1" in Fig.~\ref{Breakdown03091998} (b)); empirical breakdown probability was found from a study of   traffic data in which traffic breakdown
was observed   on 56 different days. (b) Probability of traffic breakdown at off-ramp bottleneck   
  (labeled by $\lq\lq$off-ramp bottleneck"   in Fig.~\ref{Breakdown03091998} (b));  empirical breakdown probability was found from a study of   traffic data in which traffic breakdown
was observed  on 89 different days.  
 \label{Probability_Koller} } 
\end{center}
\end{figure}
 
However, the wave duration in the data is usually less than 5 minutes
(Figs.~\ref{FreeWaves1996}--\ref{Nuclei1996_2}, \ref{Nuclei1998} and~\ref{Nuclei1998_2}).
Therefore, in    5-minutes average traffic data studied in~\cite{Brilon310,Brilon210,Brilon,BrilonISTTT2009,Elefteriadou2014}
the waves cannot usually be resolved.

We study the flow rate functions of the empirical probability of spontaneous traffic breakdown
whose nucleation is associated with
  wave propagation through highway bottlenecks.
  To find the empirical breakdown probability (black points in Fig.~\ref{Probability_Koller}), we study   data sets of 1-min averaged data measured during  in 1996--2014. In each of the data sets
 traffic breakdown has been observed. The data sets have been measured
  on the same section of the freeway A5-South as the   data   studied  above 
(Figs.~\ref{FreeWaves1996}--\ref{Nuclei1998_2}). 

Empirical
breakdown probabilities  (black points in Fig.~\ref{Probability_Koller}) are found  as   functions of the total flow rate across the road
as follows: (i)  The breakdown is measured at detector   with the use of 1-min averaged data. In the most of the data sets used for the calculation of the empirical breakdown probability
(black points in  Fig.~\ref{Probability_Koller}), a nucleus  for traffic breakdown appears during  the propagation of
one of the waves through the bottleneck  location. (ii)
 The flow rates in free flow (before the breakdown) have been averaged over 15 min intervals. (iii) The flow rate axis is divided in flow rate
 intervals $(q_{k}, \ q_{k}+\Delta q_{k})$ with
constant  $\Delta q_{k}=$ 940 vehicles/h ($\lq\lq$$k$-flow rate interval"), $k=1,2,...K$, where $K$ is the total number of different $k$-flow rate intervals in free flow; (iv) for each of the $k$-flow rate intervals,
breakdown probability is equal to $n_{k}/N_{k}$,  where  $N_{k}$ is  the number of observed flow rates within the $k$-flow rate interval in all data sets,
 $n_{k}$ is  the number of breakdowns found in the  $k$-flow rate interval.
  
We  
 have found that the empirical probabilities of traffic breakdown measured at   detector 
 as function of the flow rate (black points  in Fig.~\ref{Probability_Koller}) for both the on-ramp and off-ramp bottlenecks   
are well fitted with  a theoretical one given by formula (\ref{Prob_For}) with fitting
parameters   $(q_{\rm p}, \ \alpha^{-1})=$ (6800, 456) vehicles/h for the on-ramp bottleneck (Fig.~\ref{Probability_Koller} (a))   and
  (6600, 643) vehicles/h for the off-ramp bottleneck (Fig.~\ref{Probability_Koller} (b)).

  \subsection{Empirical permanent disturbances at highway bottlenecks and  nucleation of empirical traffic breakdown \label{Perm_Cr_Dis_S}} 

\begin{figure}
\begin{center}
\includegraphics*[width=10 cm]{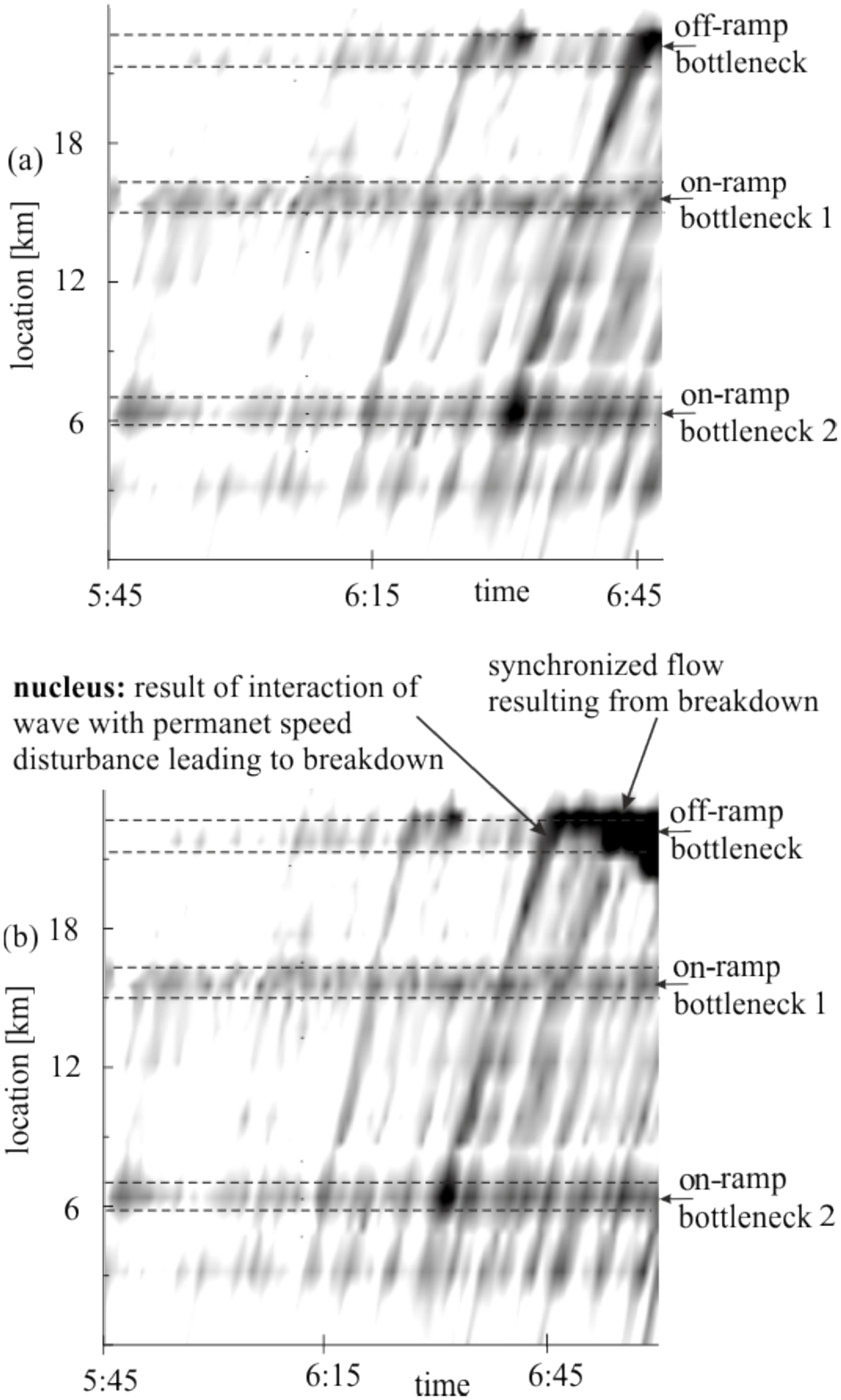}
\end{center}
\caption[]{Explanation of  physics of nuclei for traffic breakdown with empirical data:
(a) Empirical permanent local speed disturbances in free flow at highway bottlenecks for the data set
for which traffic breakdown is shown in Fig.~\ref{Nuclei1998_2}   
  (real field traffic data    measured on September 03, 1998).  (b) The same data as in (a), however, for a longer time interval showing
  that the nucleus for the breakdown at the off-ramp bottleneck
   appears due to some  interaction of
  the wave with a permanent speed disturbance at the bottleneck. In (a, b), 
  empirical data for the speed $v(x, t)$ presented by regions with variable shades
of gray; in white regions $v\geq$     115 km/h, 
  in  black  regions $v\leq$   80 km/h.  
 Parameters of the speed presentation   $v(x, t)$ made with   (\ref{var_traffic_w}) are  the same as those in 
 Fig.~\ref{FreeWaves1996}.  Narrow road regions of a smaller speed (permanent local speed disturbances), which are localized in   neighborhoods of the effective locations of 
the  bottlenecks, are marked by double dashed lines. Off-ramp bottleneck,   on-ramp bottleneck 1, and   on-ramp bottleneck 2 are the same as those in
Fig.~\ref{Nuclei1998}.
}
\label{Nuclei1998_det}
\end{figure}

In   empirical data sets, a wave moving in free flow at the velocity (\ref{wave_Vel}) acts as a nucleus for traffic breakdown
{\it only} at some   effective location of a highway bottleneck: No traffic breakdown has been observed between the bottleneck locations.
To understand this empirical result, rather than waves of $\Delta   v_{\rm wave}$ (Fig.~\ref{Nuclei1998}),
we consider empirical waves of the speed $v(x, t)$ averaged across the road (Fig.~\ref{Nuclei1998_det} (a)).

We see that additionally to waves of the speed propagating downstream,
 there are three narrow road regions, which are localized in   neighborhoods of the   locations of 
  off-ramp bottleneck,  on-ramp bottleneck 1, and   on-ramp bottleneck 2, respectively. Within these narrow regions,
  the speed is smaller than outside
them (Fig.~\ref{Nuclei1998_det} (a)). These narrow regions
of the decrease in the speed at the effective locations of the bottlenecks can be called permanent empirical local speed
disturbances in free flow at highway bottlenecks~\cite{Permanent}. 

Empirical observations show that a wave acts as a nucleus for traffic breakdown only when the wave reaches the location of a permanent local speed
disturbance  in free flow at  a highway bottleneck. For this reason, the location of the permanent   disturbance determines 
 the effective location of the bottleneck at which traffic breakdown occurs. A decrease in the free flow speed
 within   the   permanent local speed disturbance becomes larger, when the wave 
 reaches the effective bottleneck location.
 This is because within the wave the flow rate is larger and the speed is smaller than outside the wave.

 \begin{figure*}
\begin{center}
\includegraphics*[width=14 cm]{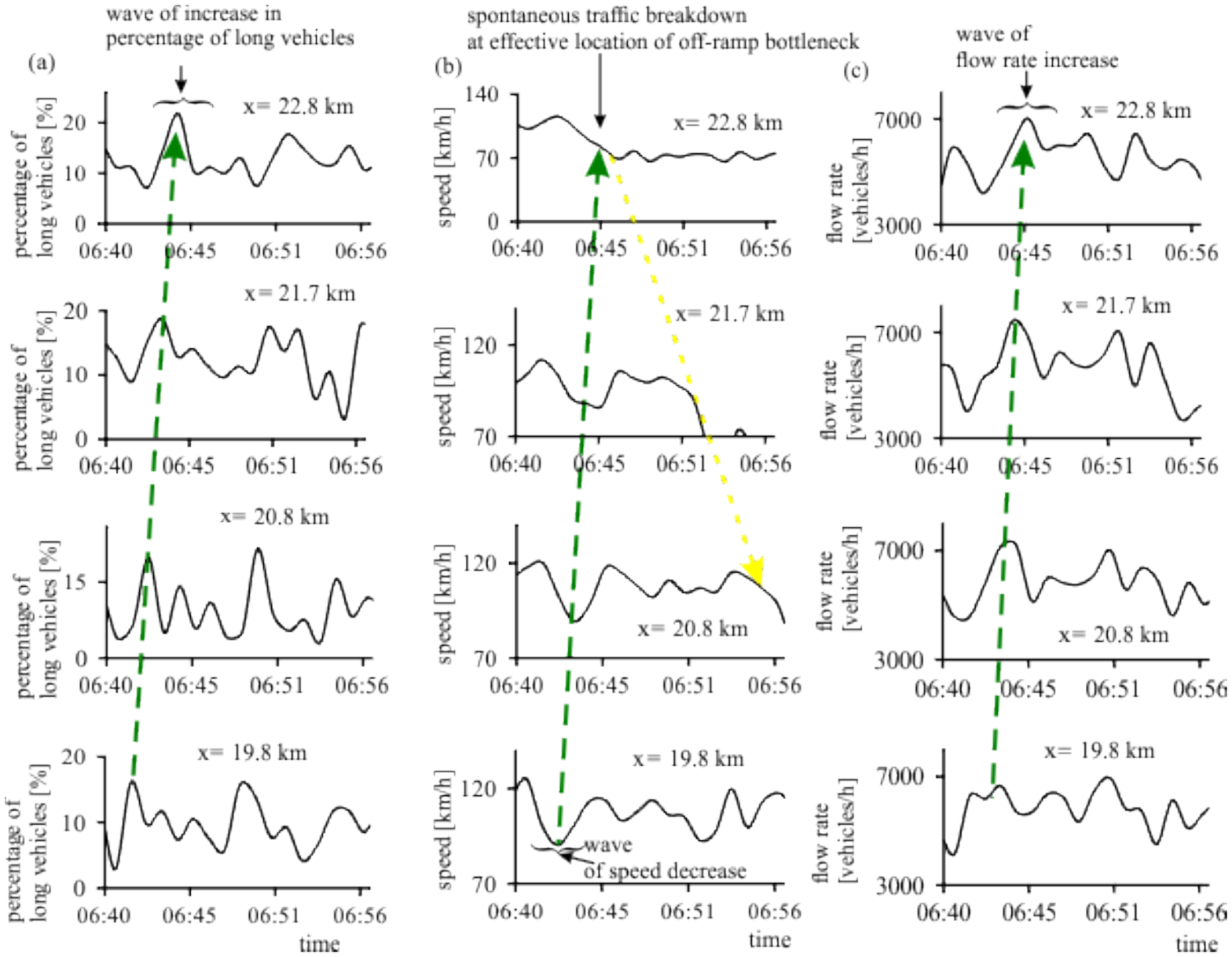}
\end{center}
\caption[]{Empirical time-distributions of traffic variables averaged across the road at different   detector locations
within a wave   that initiates the breakdown at the location of a permanent speed disturbance (effective location of the off-ramp bottleneck): (a) The percentage of long vehicles.
(b) The   speed. (c) The total flow rate. Arrows in downstream direction show 
  regions of the downstream propagation of the wave. The arrow in upstream direction in (b) shows
  the propagation of synchronized flow that has occurred due to the breakdown at the off-ramp bottleneck.
 The same real field traffic data as that shown in Fig.~\ref{Nuclei1998_2}.   
}
\label{Wellen_Diag_LKW_Con_1998_09_03}
\end{figure*}

 \begin{figure*}
\begin{center}
\includegraphics*[width=14 cm]{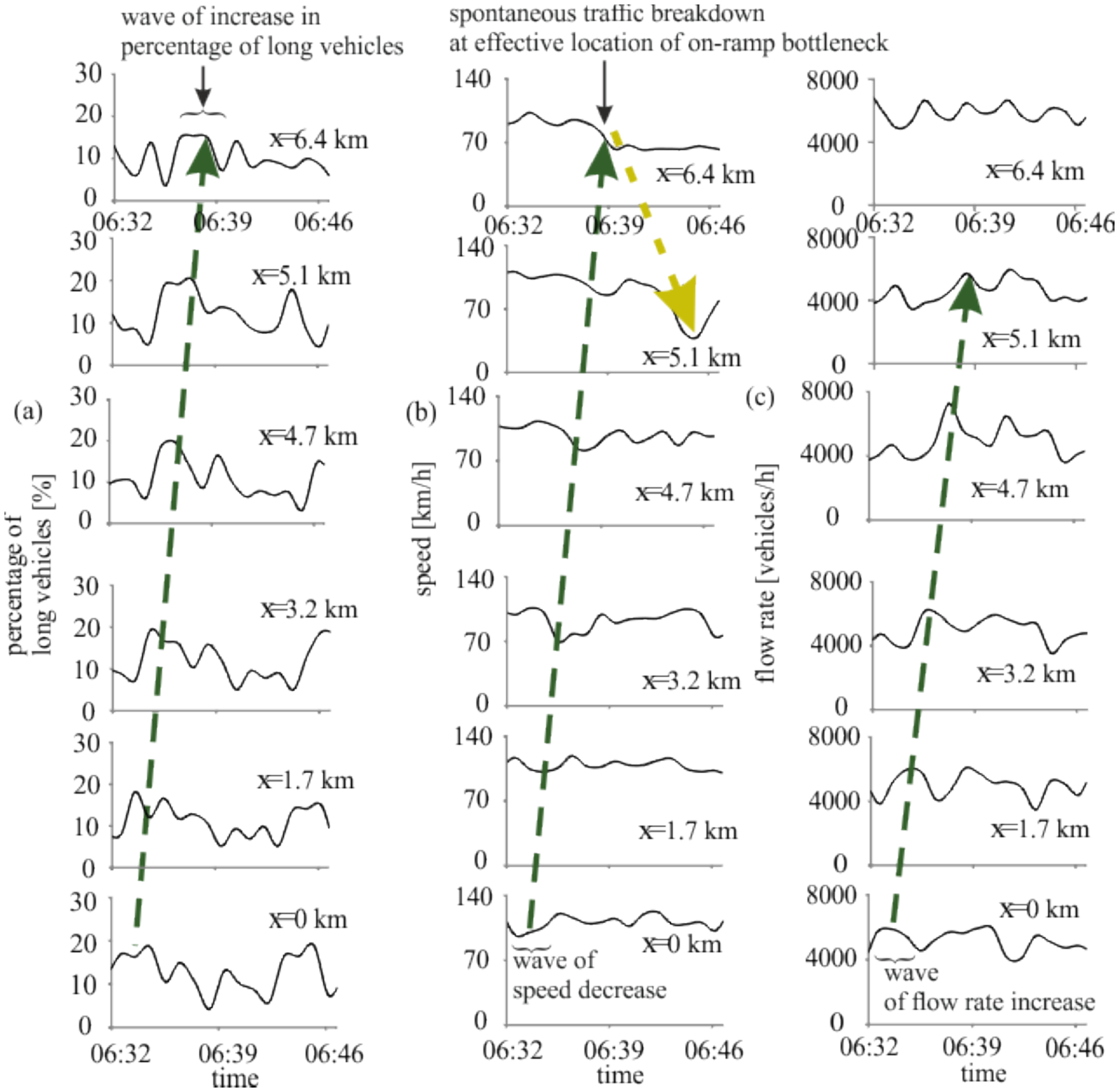}
\end{center}
\caption[]{Empirical time-distributions of traffic variables at different road detector locations
within a wave     that initiates the breakdown at the on-ramp bottleneck: (a) The percentage of long vehicles.
(b) The speed averaged across the road. (c) The total flow rate. Arrows in downstream direction show 
  regions of the downstream propagation of the wave. The arrow in upstream direction in (b) shows
  the propagation of synchronized flow that has occurred due to the breakdown at the bottleneck.
  The same real field traffic data as that shown in Figs.~\ref{Breakdown} (b) and~\ref{Nuclei1996_2}    measured on April 15, 1996; the on-ramp bottleneck   is 
the bottleneck labeled by $\lq\lq$on-ramp bottleneck 2" in Fig.~\ref{Nuclei1998_det}; the effective location
of this on-ramp bottleneck is approximately equal to $x=$ 6.4 km.}
\label{Nuclei1996_3}
\end{figure*}

Thus the physics of the occurrence of empirical nuclei for traffic breakdown at highway bottlenecks can be explained
by an interaction of a wave in free flow with a permanent   speed disturbance localized at 
 the effective location of the bottleneck.
 
 Empirical results presented in Figs.~\ref{Nuclei1998_det} (b) and~\ref{Wellen_Diag_LKW_Con_1998_09_03} for the off-ramp bottleneck
   confirm the above conclusion that a wave becomes to be nucleus for traffic breakdown {\it only} at
the effective locations of the bottleneck at which permanent   speed disturbance is localized
(Fig.~\ref{Nuclei1998_det} (a))~\cite{Helbing}. The same conclusion is valid for the on-ramp bottleneck (Fig.~\ref{Nuclei1996_3})
that is the bottleneck labeled by $\lq\lq$on-ramp bottleneck 2" in Fig.~\ref{Nuclei1998_det}.
For both off- and on-ramp bottlenecks, a wave, within which
the percentage of long vehicles (Figs.~\ref{Wellen_Diag_LKW_Con_1998_09_03} (a) and~\ref{Nuclei1996_3}(a)) 
and the flow rate are larger (Figs.~\ref{Wellen_Diag_LKW_Con_1998_09_03} (c) and~\ref{Nuclei1996_3}(c)) whereas the average speed is lower
(Figs.~\ref{Wellen_Diag_LKW_Con_1998_09_03} (b) and~\ref{Nuclei1996_3}(b)) than outside the wave, becomes to be a nucleus for  traffic breakdown only at
the effective location of the bottleneck at which a permanent   speed disturbance is localized
(Fig.~\ref{Nuclei1998_det} (a)).

The empirical evidence of   the effect of   permanent local speed
disturbances  in free flow at the effective locations of   highway bottlenecks on the breakdown nucleation due to
wave propagation   revealed in the article
confirms the theoretical explanation of an
F$\rightarrow$S transition made in   three-phase theory. In this theory,
the assumption about the existence of   permanent local speed
disturbances  in free flow at the effective locations of   highway bottlenecks should explain why the probability of the F$\rightarrow$S transition
in metastable free flow is considerably larger at the bottlenecks than outside them~\cite{KernerBook,Kerner2000A3,KKl_B13,KKl_B23,KKl2008A3}.

 More than 160 traffic breakdowns at on- and off-ramp bottlenecks on different highways in Germany that measured during 1996--2014 have been studied.
It turns out that the empirical  result of breakdown nucleation at a highway bottleneck 
due to the interaction of a wave in free flow with a permanent   speed disturbance localized at 
 the effective location of the bottleneck
is   the common one
for the most of the data sets.

 \section{Empirical spatiotemporal  structure of nuclei   and microscopic theory
 of breakdown nucleation \label{2D_Theory_S}}

 \subsection{Empirical two-dimensional (2D) asymmetric spatiotemporal structure of nuclei for traffic breakdown \label{2D_Wave_S}}

\begin{figure}
\begin{center}
\includegraphics*[width=10 cm]{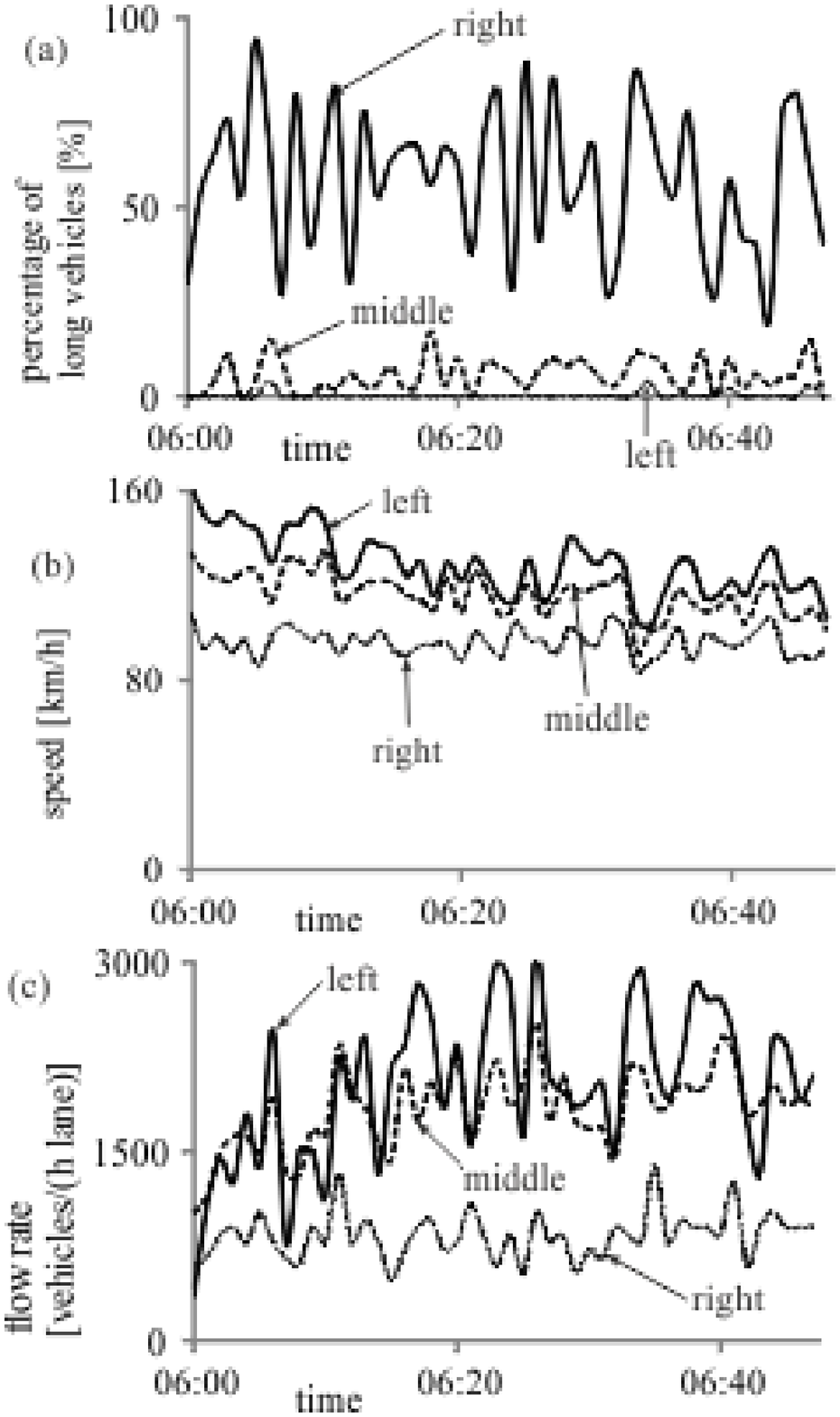}
\end{center}
\caption[]{Empirical time-dependencies of percentage of long vehicles  (a), speed (b), and flow rate (c) in different highway lanes
at location $x=$ 0 km for  real field traffic  data  shown   in Fig.~\ref{Breakdown} (b). 
}
\label{Detector1_1996}
\end{figure}

\begin{figure*}
\begin{center}
\includegraphics*[width=14 cm]{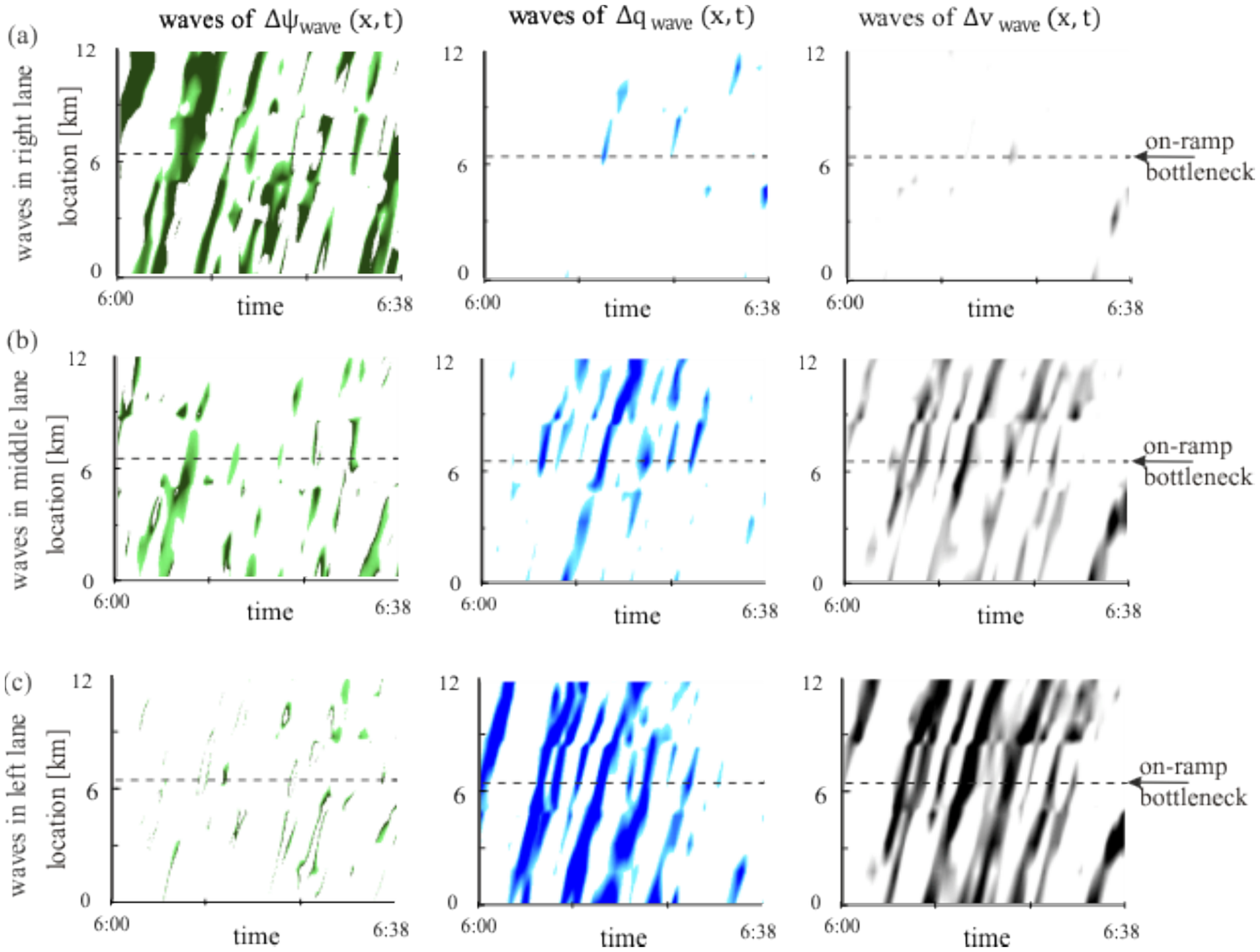}
\end{center}
\caption[]{Empirical waves of the percentage of slow (long) vehicles $\Delta \psi_{\rm wave}$  (left column),
the flow rate $\Delta   q_{\rm wave}$  (middle column), and the vehicle speed $\Delta   v_{\rm wave}$  (right column)
  in free flow  for different road lanes for real field traffic  data  shown in Fig.~\ref{Breakdown} (b)
during the same time interval as that in Fig.~\ref{FreeWaves1996}:  (a)  Right lane.
 (b) Middle lane. (c) Left lane.   Waves of
  $\Delta \psi_{\rm wave}(x, t)$ are presented  by regions with 
variable shades of gray (green   in the on-line version)  (in white regions $\Delta \psi_{\rm wave}\leq$ 0.3 $\%$,
in black (dark green) regions $\Delta \psi_{\rm wave}\geq$ 1 $\%$).  
Waves of   $\Delta q_{\rm wave}(x, t)$   are
 presented  by regions with 
variable shades of gray (blue   in the on-line version)  (in white regions $\Delta q_{\rm wave}\leq$ 480 vehicles/h,
in black (dark blue) regions $\Delta q_{\rm wave}\geq$ 800 vehicles/h).  
Waves of   $\Delta v_{\rm wave}(x, t)$   are presented  by regions with 
variable shades of gray    (in white regions $\Delta v_{\rm wave}\leq$ 7 km/h,
in black  regions $\Delta v_{\rm wave} \geq$ 20  km/h).
Model parameters in formula (\ref{var_traffic_w}) are the same as those in Fig.~\ref{FreeWaves1996}.
}
\label{Nuclei1996_lane}
\end{figure*}

To study a possible effect of a non-homogeneity of traffic flow across the road on the breakdown nucleation, we consider
empirical traffic variables in different freeway lanes (Figs.~\ref{Detector1_1996}
and~\ref{Nuclei1996_lane}).   We should mention that the most of long vehicles move
in the right lane (sometimes traffic flow in the right lane consists of almost 100$\%$ (slow) long vehicles) (Fig.~\ref{Detector1_1996} (a)).
The percentage of long vehicles $\psi$ in the middle lane is considerably smaller than in the right lane;
almost no long vehicles   move in the left lane (Fig.~\ref{Detector1_1996} (a)).

We have found the following  empirical result:
An empirical   wave  in free flow 
 exhibits   
a  two-dimensional (2D)  structure: Wave's characteristics 
 are different in different highway lanes (Fig.~\ref{Nuclei1996_lane}). This wave structure is   asymmetric  
for different traffic variables in the perpendicularly  direction to the flow direction.
The most waves of long vehicles are observed in the right lane, while in the left lane
almost no waves of $\Delta \psi_{\rm wave}$ exist. On contrary, the most waves of the flow rate and vehicle speed
are observed in the left lane, while in the right lane
almost no waves of $\Delta q_{\rm wave}$ and $\Delta v_{\rm wave}$ exist.

  However, to understand the effect of 2D asymmetric structure of nuclei on features of traffic breakdown,  
  a study of   
  {\it microscopic} empirical
  data  is required in which lane changing  and vehicle acceleration (deceleration) in a neighborhood of a bottleneck can be resolved.
  Unfortunately, such   microscopic (single-vehicle) empirical data for free flow at bottlenecks is currently not available.
  Therefore, in Sec.~\ref{Phys_S1} with the use of a three-phase stochastic microscopic traffic flow model,
  we study  theoretical predictions   about the effect of 2D asymmetric structure of nuclei
  on microscopic features of   traffic breakdown at an on-ramp bottleneck.
 
 \subsection{Microscopic theory
 of the nucleation of traffic breakdown at highway bottlenecks \label{Phys_S1}}
 
 \subsubsection{Model \label{Phys_Mod_S1}}

 We consider a  simple  model of traffic flow on two-lane road with an on-ramp bottleneck. In this model, 
   we assume that traffic flow consists 
   of identical passenger vehicles in which there is only one
   slow vehicle   moving in the right road lane.
   Such a model of traffic flow with a slow vehicle is known as a $\lq\lq$moving bottleneck" 
   model~\cite{Gazis1992,Newell1993,Newell1988,Daganzo2002,Lebacque1998,Leclercq,Daganzo2003,Rakha2014A,Rakha2014B,KKl2010A}.
   
\begin{figure}
\begin{center}
\includegraphics*[width=9 cm]{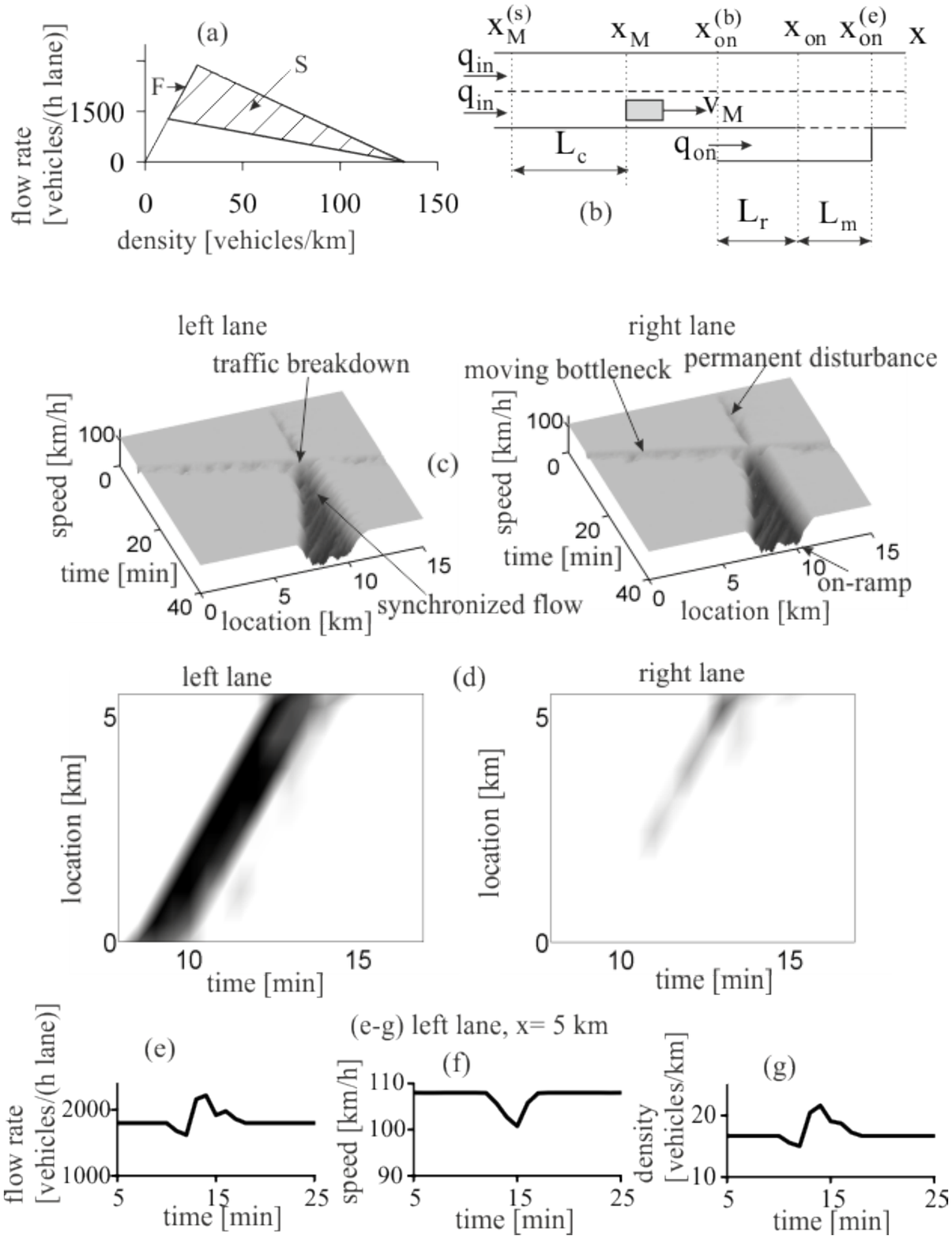}
\caption{Simulations of the nucleation of traffic breakdown on two-lane road with on-ramp bottleneck through  wave propagation in free flow:
(a) Hypothesis of three-phase theory
about 2D-steady states of synchronized flow   incorporated in a stochastic microscopic three-phase model used for simulations.
(b) Models of moving and on-ramp bottlenecks.
(c) Nucleation of traffic breakdown at on-ramp bottleneck through  wave propagation caused by a moving bottleneck in different road lanes (left -- left lane, right -- right lane).
(d) Wave of the flow rate 
     $q (x, t)$ in different road lanes  (left -- left lane, right -- right lane) 
 presented  by regions with 
variable shades of gray    (in white regions $q  \leq$ 2000 vehicles/h,
in black  regions $q \geq$ 2150 vehicles/h);  
   wave presentation is made with the use of  a virtual detector moving at the speed
 $v_{\rm M}$ (Appendix~\ref{Sim_MB_S}). 
 (e--g) Time-functions of the flow rate (e), the speed (f) and the density (g) within the wave in the left lane at location $x=$ 5 km. $v_{\rm M}=$ 82.8 km/h,
 $(q_{\rm in}, \ q_{\rm on})=$ (1800, 750) vehicles/h, $x_{\rm on}=$ 10 km.
 Other model parameters are in Table~\ref{table2} of Appendix~\ref{D_Model}.
 \label{Model} } 
\end{center}
\end{figure}

   However, as explained in details in~\cite{Kerner_Review},
    traffic flow models used 
   in~\cite{Gazis1992,Newell1993,Newell1988,Daganzo2002,Lebacque1998,Leclercq,Daganzo2003,Rakha2014A,Rakha2014B} cannot describe
 an  F$\rightarrow$S transition in metastable free flow at a highway bottleneck,
   as  observed in all known measured traffic data~\cite{KernerBook} including
   empirical field traffic  data studied in this article. Therefore, we make simulations
   with a stochastic microscopic three-phase traffic flow model with on-ramp and moving bottlenecks
  of Ref.~\cite{KKl2010A}. In this model,   states of synchronized flow
  cover a two-dimensional region (2D) in the flow--density plane (Fig.~\ref{Model} (a)), as it follows from a hypothesis of three-phase theory about  
   states of synchronized flow~\cite{Kerner1998A,Kerner1999A,2D}. 
  As shown in~\cite{KKl2010A,KKl,KKl2003A,Kerner2008,KKl2009A}, this model can reproduce all known empirical macroscopic features
  of traffic breakdown (F$\rightarrow$S transition) in metastable free flow at  highway bottlenecks~\cite{M_3}.

A moving bottleneck is caused by a slow vehicle that moves in the right lane at a maximum speed $v_{\rm M}$
that is lower than the   speed of other (identical)
  vehicles  
$v_{\rm free}$. It is assumed  that there is a road region with length $L_{\rm c}$
upstream of the slow vehicle
(Fig.~\ref{Model} (b)). This  region is a {\it moving} one  at 
the   speed   $v_{\rm M}$.  
 Within this moving   region  $L_{\rm c}$,   called   as $\lq\lq$moving 
 merging region" of the moving bottleneck,  all passenger vehicles moving in the right lane 
 change to the left lane, while approaching the moving bottleneck; this lane changing occurs      independent of the speed difference between lanes,
  when some safety  
 conditions are satisfied (Sec.~\ref{Model_S_M}). 
The   length   $L_{\rm c}$ of the
moving 
 merging region is associated with the mean distance at which vehicles
 recognize
the slow vehicle.  
In accordance with~\cite{KKl2010A}, a moving bottleneck
can be a nucleus for traffic breakdown (F$\rightarrow$S transition)
at an on-ramp bottleneck (Fig.~\ref{Model} (c)).
Because  the rules of vehicular motion in the
 three-phase traffic flow model as well as in the bottleneck models (Fig.~\ref{Model} (b)) have
 been presented and discussed in details in~\cite{KKl2010A}, they are given in Appendix~\ref{D_Model}.

    \begin{figure}
\begin{center}
\includegraphics*[width=9 cm]{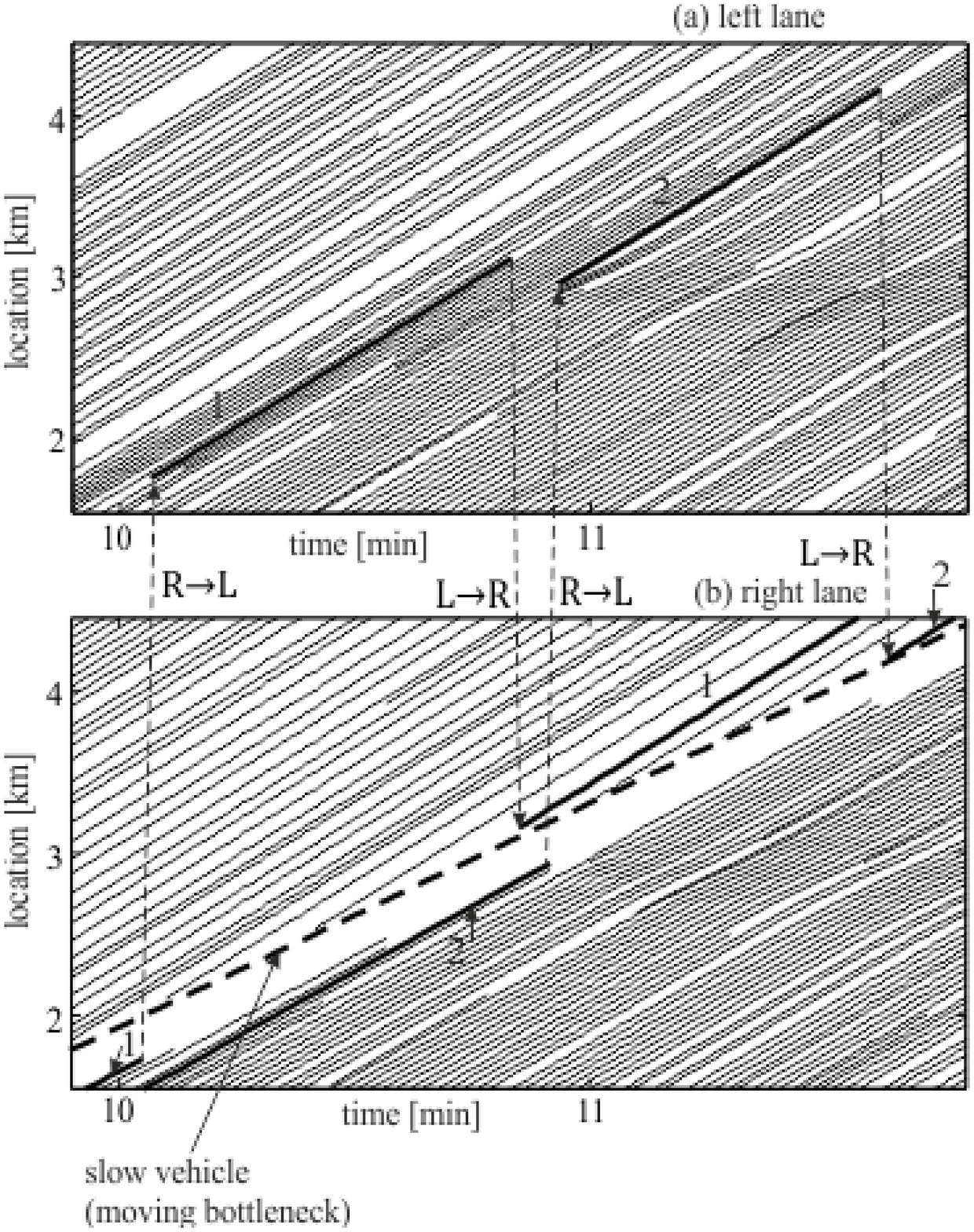}
\caption{Fragment of simulated vehicle trajectories of sequence R$\rightarrow$L$\rightarrow$R of lane changing in a neighborhood of the slow vehicle (moving bottleneck)
associated with simulations shown in Fig.~\ref{Model} (c, d):
(a)  Left lane. (b) Right lane.
\label{Theory2} } 
\end{center}
\end{figure}

\subsubsection{Microscopic 2D asymmetric structure of waves in free flow \label{Phys_W_S1}}

To understand   the 2D asymmetric structure of empirical waves in free flow (Sec.~\ref{2D_Wave_S})  
as well as  a possible impact of this   wave structure
 on   traffic breakdown,
in comparison with~\cite{KKl2010A}, we   study here the effect of the moving bottleneck on  spatiotemporal distributions of
traffic variables  in different road lanes  (Fig.~\ref{Model} (d--g)), 
  on  lane changing behavior in a neighborhood of the moving bottleneck upstream  of the on-ramp bottleneck
(Fig.~\ref{Theory2})
and in a neighborhood of the on-ramp bottleneck (Fig.~\ref{Theory3}).

To  pass  the slow vehicle (moving bottleneck), vehicles moving in the right lane change to the left lane within the merging region of the moving bottleneck
(up-arrows R$\rightarrow$L for vehicles 1 and 2 in Fig.~\ref{Theory2}). After passing the slow vehicle, most of these vehicle change back to the right lane  
(down-arrows L$\rightarrow$R for vehicles 1 and 2 in Fig.~\ref{Theory2}). This sequence R$\rightarrow$L$\rightarrow$R of lane changing leads to an increase in the flow rate
in the left lane 
in a neighborhood of the slow vehicle  (Fig.~\ref{Model} (d)). Therefore, this increase in the flow rate in the left lane
moves with the speed of the slow vehicle: A wave of the increase in the flow rate
 occurs that moves at the velocity $v_{\rm M}$ (Fig.~\ref{Model} (d)). Due to the increase in the flow rate,
 the speed in the left lane within the wave decreases. Time-distributions of the flow rate, speed, and density
within the wave in the left lane (Fig.~\ref{Model} (e--g)) confirm these conclusions about the wave features.

There is also an increase in the flow rate in the right lane. 
This increase in the flow rate is due to an increase in the vehicle density   upstream of the moving bottleneck (Fig.~\ref{Model} (b)); however, this increase in the flow rate
is considerably smaller than that in the left lane. 
Thus the wave of the flow rate exhibits
a 2D asymmetric    structure   whose characteristics 
 are different in different highway lanes (Fig.~\ref{Model} (d)).
 This is qualitatively exactly the same effect as observed in real field empirical data
 (Sec.~\ref{2D_Wave_S}).
 
These simulations   allow us to explain empirical   waves in free flow (Secs.~\ref{Wave_Sub_S} and~\ref{2D_Wave_S}) as follows.
In empirical data,  the percentage of slow (long) vehicles in the right lane exhibits   large
oscillations over time
(Fig.~\ref{Detector1_1996}(a)). Therefore, waves of slow   vehicles occur in the right lane.  
Because the most of   slow vehicles move in the right lane  (Fig.~\ref{Detector1_1996} (a)), 
the largest wave amplitude $\Delta \psi_{\rm wave}$ is observed in the right lane  (left column in Fig.~\ref{Nuclei1996_lane} (a)),
whereas  almost no waves of slow vehicles is observed in the left lane (left column in Fig.~\ref{Nuclei1996_lane} (c))~\cite{Middle}.

There are very different empirical   vehicle speeds in the left and   middle lanes  (Fig.~\ref{Detector1_1996}(b)). This allows us to assume that 
there are passenger vehicles that prefer the uninterrupted  motion at a   higher speed in the left lane, and 
those passenger vehicles that prefer the motion   in the middle  and right lanes at a lower speed, while using the left lane for passing only.

We can assume that the frequency of
this passing increases considerably when      passenger vehicles 
  approach a   wave   of slow vehicles in the right lane: The passenger vehicles change firstly to the left lane
  and, after the vehicles have   passed the wave, they change back to the middle    and to the right lanes. This is qualitatively the same effect as  
   the sequence R$\rightarrow$L$\rightarrow$R of lane changing in the neighborhood of the slow vehicle found in simulations (Fig.~\ref{Theory2}).
 
  Thus, during the passing of a wave of slow vehicles in the right lane,
  the flow rate in the left lane  increases firstly, and then the flow rate decreases. As in simulations (Fig.~\ref{Model} (d)), this should  lead to the wave of the flow rate in the left lane
 caused by the wave of   slow vehicles in the right lane (Fig.~\ref{Nuclei1996_lane}).  
 This explains the emergence of  waves
 $\Delta q_{\rm wave}$ moving at the average speed of slow vehicles
 (\ref{wave_Vel}) (Figs.~\ref{Nuclei1996} and~\ref{Nuclei1996_lane})
 that is considerably smaller than the vehicle speed in the left and   middle lanes (Fig.~\ref{Detector1_1996}).
 As in simulations,   the increase of the flow rate
 in free flow leads to a speed decrease; therefore, as in simulations (Figs.~\ref{Model} (e, f)), in empirical observations
 a 2D-structure of the waves of the speed and   flow rate coincide qualitatively each other (middle and right columns in Fig.~\ref{Nuclei1996_lane}).

 \subsubsection{Microscopic structure of permanent speed disturbance at  
 bottleneck \label{Phys_D_S2}}
 
\begin{figure}
\begin{center}
\includegraphics*[width=9 cm]{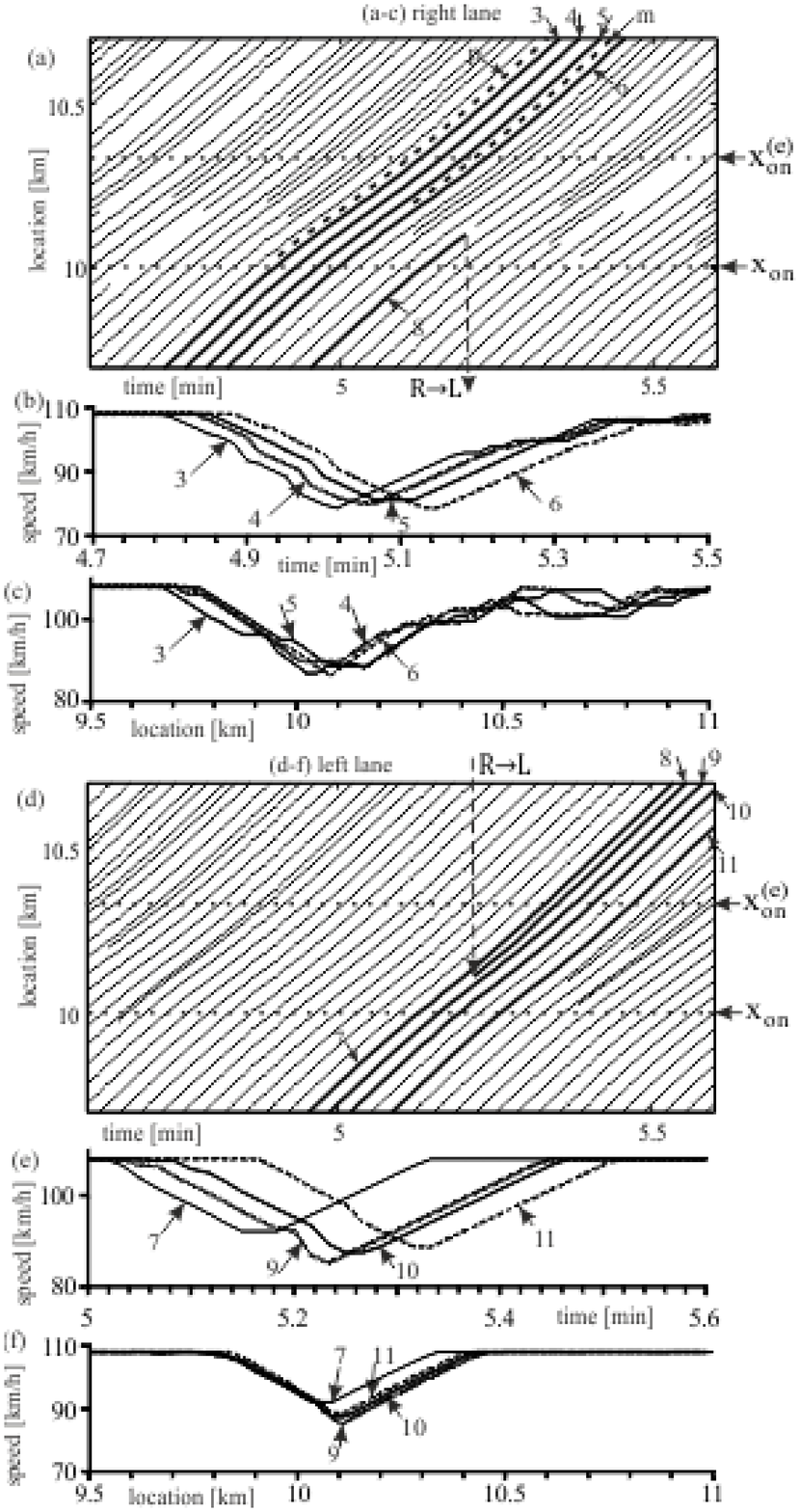}
\caption{Simulations of   permanent local disturbance in free flow at on-ramp bottleneck associated with  Fig.~\ref{Model} (c):
(a--c) Fragment of vehicle trajectories (a) and microscopic (single-vehicle) speed along chosen trajectories (b, c) in time (b) and location (c)
whose numbers are the same in (a) and (b, c)
 in the right lane. (d--f) Fragment of vehicle trajectories (d) and microscopic speed along chosen trajectories (e, f) in time (e) and location (f) whose numbers are the same in (d) and (e, f)
 in the left lane. 
 \label{Theory3} } 
\end{center}
\end{figure}

 In simulations, there is a permanent speed disturbance at the on-ramp
 bottleneck (Fig.~\ref{Theory3}). To explain the disturbance, note  when the flow rate $q_{\rm in}$ (Fig.~\ref{Model} (b)) is   large enough, 
 due to   vehicle merging from the on-ramp into the right lane of the main road (vehicle $\lq\lq$p" in Fig.~\ref{Theory3} (a)),
 the following vehicle moving in the right lane on the main road decelerates (vehicle 3 in Fig.~\ref{Theory3} (a--c)).
 This deceleration of vehicle 3 forces the following vehicles 4 and 5 to decelerate:
 The speed disturbance occurs in a neighborhood of the merging region
 on the on-ramp ($x_{\rm on} \leq x \leq x^{\rm (e)}_{\rm on}$ in Figs.~\ref{Model} (b) and~\ref{Theory3} (a--c)).
 Although the minimum speed within the disturbance increases over time (vehicle 3--5 in Fig.~\ref{Theory3} (b, c)), the disturbance
 is maintained on average because   next vehicle that merges from the on-ramp onto the main road (vehicle $\lq\lq$m" in Fig.~\ref{Theory3} (a))
 leads to the deceleration of the following vehicle (vehicle 6 in Fig.~\ref{Theory3} (a--c)), and so on.

The permanent speed disturbance occurs also in the left lane
(vehicle 7 in Fig.~\ref{Theory3} (d--f)). The disturbance appears due to  vehicles that decelerate initially
within the disturbance in the right lane and then they change to the left lane as shown with an example of 
  vehicle 8 together with arrow labeled by R$\rightarrow$L in Fig.~\ref{Theory3} (a, d).
  Due to this lane changing, the following vehicle 9 must decelerate to a smaller speed than
  that  of vehicle 8. The lane changing maintains the permanent speed disturbance  in the left lane
  (vehicles 9--11 in Fig.~\ref{Theory3} (d--f)).

   \subsubsection{Microscopic features of interaction of  waves in free flow with permanent local disturbance
   at bottleneck \label{Phys_In_S2}}

   \begin{figure}
\begin{center}
\includegraphics*[width=9 cm]{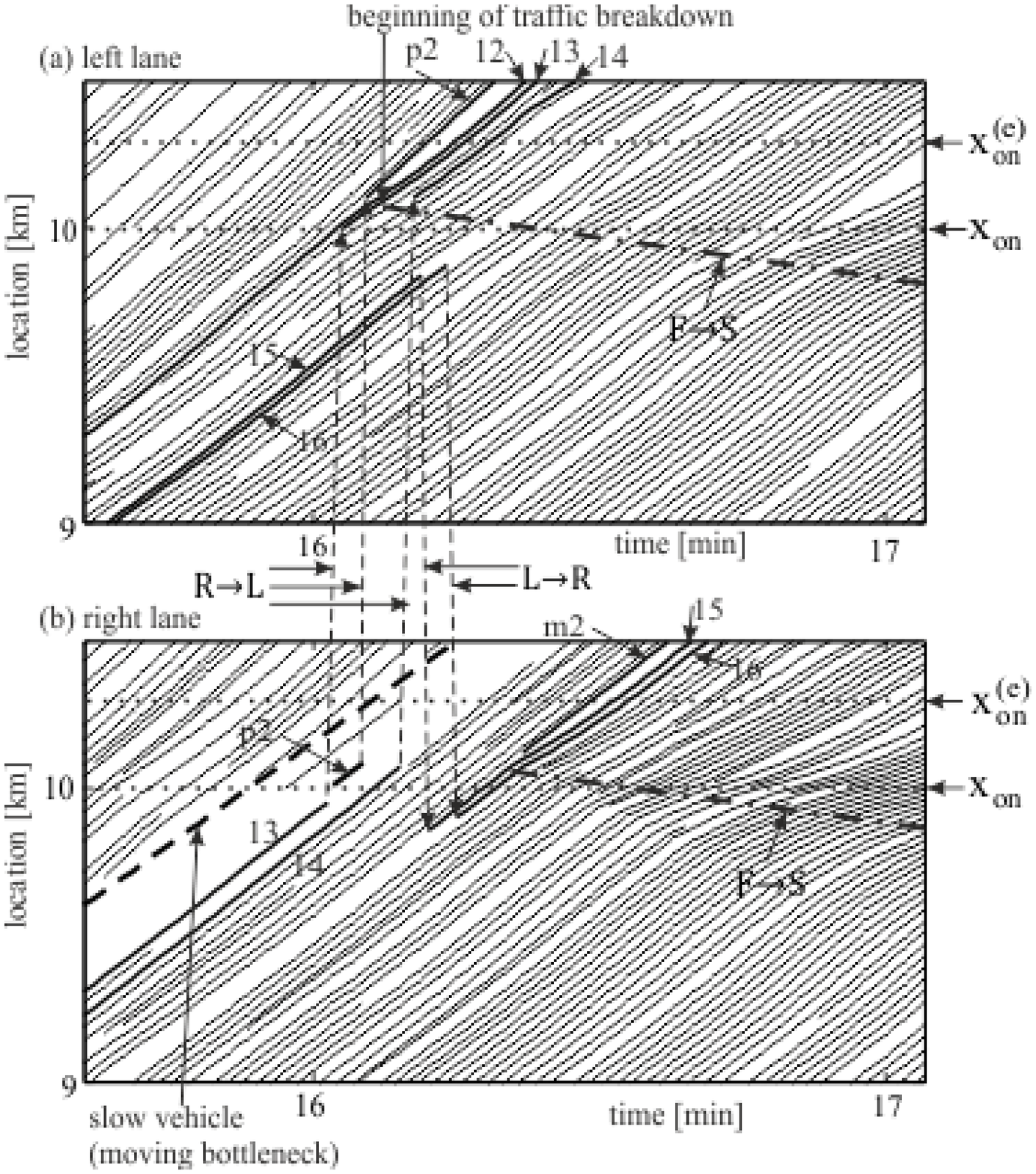}
\caption{Fragments of simulated vehicle trajectories associated with  Fig.~\ref{Model} (c):
 (a) Left lane. (b) Right lane. 
 \label{Theory4} } 
\end{center}
\end{figure}
 
  Before the moving bottleneck reaches the on-ramp bottleneck,
 no spontaneous traffic breakdown occurs at the on-ramp regardless of the existence
 the permanent speed disturbance. This is because     at chosen flow rates $q_{\rm in}$ and $q_{\rm on}$ 
  the probability of spontaneous traffic breakdown due to model fluctuations  
  is   small (although this probability is larger than zero).
 However,   when
  the wave
 reaches the effective location of the on-ramp bottleneck (Fig.~\ref{Model} (c)), the speed decreases at the bottleneck additionally to that 
 within  the permanent speed disturbance;
as a result,  the wave becomes to be a nucleus for spontaneous traffic breakdown at the bottleneck.
 
In simulations, this traffic breakdown  is associated with the interaction of the permanent speed disturbance with the wave of the flow rate and speed
  caused by the moving bottleneck. The breakdown
 begins to develop in the left lane in which the flow rate is larger and the speed is smaller than outside the wave.
  Firstly, within the region of the permanent disturbance ($x_{\rm on} \leq x \leq x^{\rm (e)}_{\rm on}$ in Figs.~\ref{Model} (b) and~\ref{Theory4}),
  vehicle $\lq\lq$p2" merges from the on-ramp into the right lane; this vehicle changes quickly to the left lane
   (up-arrow R$\rightarrow$L for vehicle $\lq\lq$p2" in Fig.~\ref{Theory4}).
   Due to the increase in the flow rate within the wave  in the left lane, following vehicles 12--14 should decelerate.
   This vehicle deceleration causes traffic breakdown:
   the upstream front of synchronized flow is  forming
   in the left lane while propagating upstream (dotted-dashed line labeled by F$\rightarrow$S in Fig.~\ref{Theory4} (a)).
   Vehicles 15 and 16 approaching this front of synchronized flow change to the right lane 
  (down-arrows L$\rightarrow$R for vehicles 15 and 16 in Fig.~\ref{Theory4}).
 However,   vehicles 15 and 16 should follow vehicle $\lq\lq$m2" that has just merged from the on-ramp into the right lane.
 Thus vehicles 15 and 16 should decelerate. 
  This vehicle deceleration causes traffic breakdown:
   the upstream front of synchronized flow is  forming
   in the right lane while propagating upstream (dotted-dashed line labeled by F$\rightarrow$S in Fig.~\ref{Theory4} (b)).

 In contrast to simulations with   a single slow vehicle (Fig.~\ref{Model} (b)), in real field data there are many slow vehicles (Secs.~\ref{Nuclei_S} and~\ref{2D_Wave_S}).
However,  in the empirical data there are large time-oscillations of the percentage of slow vehicles    (Fig.~\ref{Detector1_1996} (a)). For this reason, a sequence of waves
of  slow vehicles occurs (left column  in Fig.~\ref{Nuclei1996_lane}).  
 Based on a simple model   with the single slow vehicle,
we have simulated one of such waves of real traffic (Figs.~\ref{Model}--\ref{Theory4}). We have
found that, as in empirical observations (Fig.~\ref{Nuclei1996_lane}), in  simulations  the wave  exhibits
  a 2D-structure   of the flow rate and   speed (Fig.~\ref{Model} (d--f)).
  Moreover, as in empirical observations (Sec.~\ref{Nuclei_S}),
  in simulations we have found that a nucleus for traffic breakdown (F$\rightarrow$S transition)
  at the bottleneck (Figs.~\ref{Model} (c) and~\ref{Theory4}) occurs due to an interaction of this
  wave with a permanent speed disturbance at a highway bottleneck.
  Thus the above simulations can (at least qualitatively) explain the   physics of
   empirical findings of Secs.~\ref{Nuclei_S} and~\ref{2D_Wave_S}.

\section{Discussion \label{Dis_S}}
 
 \subsection{Sources of nucleus for empirical   traffic breakdown  \label{Ind_Sp_FS_S}}
 
 Both an empirical wave in free flow  (Figs.~\ref{Breakdown} (b) and~\ref{Nuclei1996_2}) 
 and a localized congested pattern (wide moving jam   in Fig.~\ref{Breakdown} (c)) become to be   nuclei   for traffic breakdown, when they
 reach  the effective location of a highway bottleneck. However, the propagation of a single
  congested pattern  to the effective bottleneck location  is sufficient for the inducing of the breakdown
at the bottleneck. In contrast, many waves in free flow can propagate through the bottleneck while initiating no breakdown at 
the bottleneck (Figs.~\ref{Nuclei1996}   and~\ref{Nuclei1998}).

The latter empirical result allows us to assume that at a given flow rate in free flow at a highway bottleneck
there is a {\it critical wave}  related to a {\it critical nucleus} for traffic breakdown.  Therefore,  if a wave is a
smaller one  than the critical wave for a given flow rate at a highway bottleneck, then no breakdown occurs while the wave propagates through the bottleneck. For example, all waves
shown in Fig.~\ref{Nuclei1996} for   $t<6:35$ and in Fig.~\ref{Nuclei1998} for   $t<6:40$ should be smaller than critical waves.
However,   waves that become to be  nuclei  for the breakdown at the effective locations of the bottlenecks in
Figs.~\ref{Nuclei1996_2}  and~\ref{Nuclei1998_2} should be 
 equal to or larger ones than   critical waves for the breakdown at the related bottlenecks, respectively.

In contrast with waves in free flow,  within a congested pattern the speed is usually   smaller than a critical speed required for the breakdown in free flow
at a bottleneck.
For this reason, at the flow rate satisfying condition $q_{\rm sum}> C_{\rm min}$,  
any localized congested pattern becomes to be a nucleus for traffic breakdown, when the pattern reaches the effective 
  bottleneck location.

Thus   a basic difference between {\it empirical spontaneous} (Fig.~\ref{Breakdown} (b)) and {\it empirical induced} breakdowns (Fig.~\ref{Breakdown} (c)) is as follows:
To initiate the spontaneous breakdown at the bottleneck, i.e., to be a nucleus for the breakdown,
  a wave in free flow should be equal to or a larger one than a critical wave.
In contrast, a localized congested pattern is always a nucleus for the breakdown at the bottleneck, when
condition $q_{\rm sum}> C_{\rm min}$ is satisfied, i.e., when traffic breakdown can occur at the bottleneck.

  \begin{figure}
\begin{center}
\includegraphics*[width=9 cm]{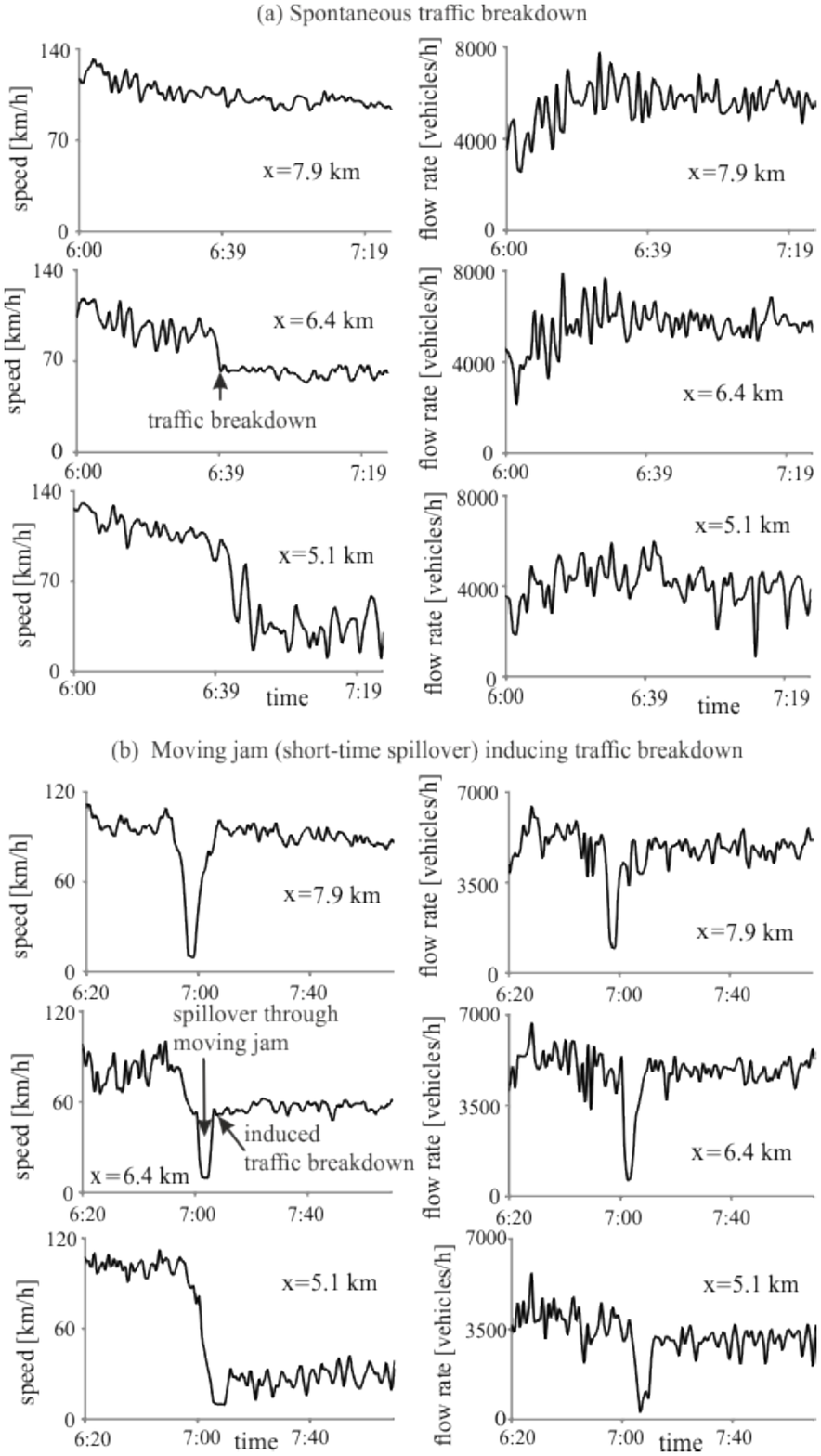}
\caption{Empirical 1-min average speed (left column) and flow rate (right column) as time-functions
measured by detectors installed on freeway A5-South at   the bottleneck location ($x=6.4$ km), downstream ($x=$ 7.9 km)
and upstream of the bottleneck ($x=$ 5.1 km): (a) Real field traffic
data measured on April 15, 1996 (Fig.~\ref{Breakdown} (b)).
(b) Real field traffic data measured on March 22, 2001 (Fig.~\ref{Breakdown} (c)). 
 \label{15041996_22032001} } 
\end{center}
\end{figure}

However, after the breakdown has occurred, characteristics of
a congested pattern that has been formed at the bottleneck do not depend on whether the congested pattern has occurred
due to   empirical spontaneous breakdown or due to   empirical induced  breakdown. This statement
is illustrated by empirical data presented in Fig.~\ref{15041996_22032001}. 
In Fig.~\ref{15041996_22032001} (a), a congested pattern at the on-ramp bottleneck (Fig.~\ref{Breakdown} (b)) has occurred due to empirical spontaneous breakdown
caused by a wave that becomes to be a nucleus for the breakdown, when the wave is at the effective bottleneck location (Fig.~\ref{Nuclei1996_2}).
In contrast,  in Fig.~\ref{15041996_22032001} (b) a congested pattern at the on-ramp bottleneck (Fig.~\ref{Breakdown} (c)) has been induced due to
the propagation of a wide moving jam through  the bottleneck. 
Empirical studies show that features of  congested traffic resulting from the induced breakdown
(at  $t>$ 7:07 in Fig.~\ref{Breakdown} (c)) 
are qualitatively identical to those found in congested traffic resulting from empirical spontaneous traffic
breakdown (Fig.~\ref{Breakdown} (b)). In particular, in both cases congested traffic resulting from the
breakdown at the bottleneck is self-maintained under free flow conditions downstream of the bottleneck.

  \begin{figure}
\begin{center}
\includegraphics*[width=9 cm]{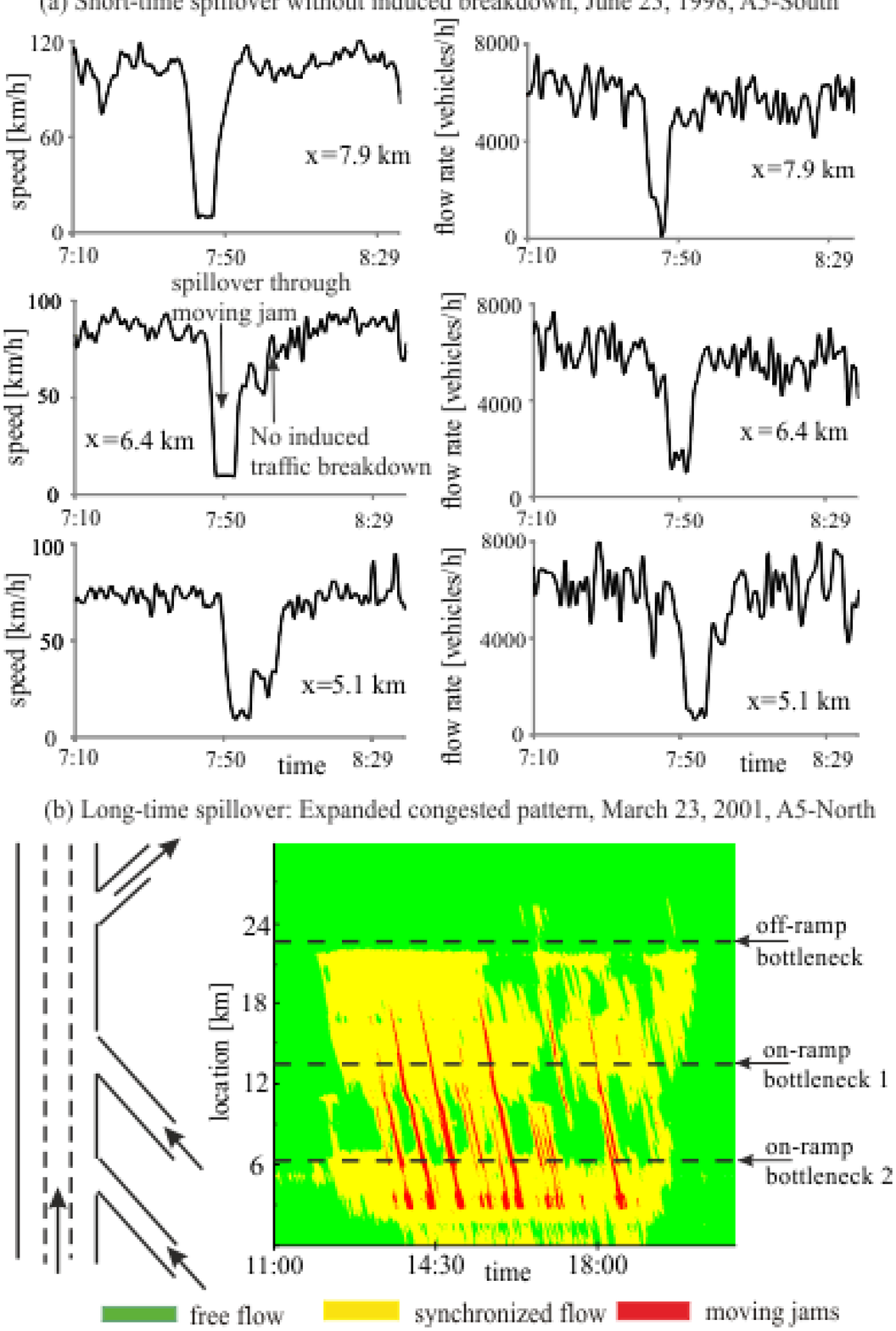}
\caption{Empirical  examples of  spillover  without induced traffic breakdown
(real field traffic data measured by road detectors on three-lane freeway A5-South (a) and A5-North (b) in Germany): (a) Short-time spillover through moving jam propagation without induced
traffic breakdown at on-ramp bottleneck. In (a), empirical 1-min average speed (left column) and flow rate (right column) as time-functions
measured by detectors installed on freeway A5-South at   the bottleneck location ($x=6.4$ km), downstream ($x=$ 7.9 km)
and upstream of the bottleneck ($x=$ 5.1 km) are shown;
data was measured on June 23, 1998 (Fig.~\ref{Breakdown} (d)).
 (b) Long-time spillover leading to expanded congested   pattern (right) measured on March 23, 2001 and scheme of freeway section of freeway A5-North
with three bottlenecks (left). Bottlenecks in (b)  have been explained   in Sec.~9.2.2
of~\cite{KernerBook}.
 \label{23061998S_23032001N} } 
\end{center}
\end{figure}

This shows that  rather than the nature of traffic breakdown,
 the  terms  {\it empirical spontaneous} and {\it empirical induced}
 traffic breakdowns  at a bottleneck
distinguish     different {\it  sources}  of  
    a  nucleus that occurrence leads to   traffic breakdown:
In Fig.~\ref{Breakdown} (b),  the source of   empirical spontaneous breakdown is
  one of the waves in free flow shown in Fig.~\ref{Nuclei1996_2}. In Fig.~\ref{Breakdown} (c),
  the source of   empirical induced breakdown is
 the wide moving jam.
 
 In contrast with the wide moving jam shown in Fig.~\ref{Breakdown} (c),
 a wide moving jam shown in Fig.~\ref{Breakdown} (d) does not induce
 traffic breakdown at the bottleneck. Indeed, in the latter case,
 after the jam is far away upstream of the bottleneck,
 free flow returns both at the effective bottleneck location as well as
 downstream and upstream of the bottleneck (Fig.~\ref{23061998S_23032001N} (a)).
 Because under condition $q_{\rm sum}> C_{\rm min}$  the jam is always a nucleus for traffic breakdown at the bottleneck,
 the case shown in Fig.~\ref{Breakdown} (d) should be related to the opposite condition
 $q_{\rm sum}< C_{\rm min}$ at which no traffic breakdown can occur at the bottleneck.
 We see that   empirical induced 
 traffic breakdown is probably the only one $\lq\lq$method" to find
 whether traffic breakdown can occur at the bottleneck or not.

 This emphasizes another difference between   empirical spontaneous  and   empirical induced 
 traffic breakdowns  at a highway bottleneck:
 When waves in free flow propagate though the bottleneck without 
 initiating of the breakdown, we cannot state whether
 all waves are smaller than a    critical wave, or
 condition $q_{\rm sum}< C_{\rm min}$ is satisfied at which no traffic breakdown can occur at the bottleneck.
 In contrast, when a local congested pattern propagates through the bottleneck
 without 
 inducing  of the breakdown, we can state that the flow rate at the bottleneck is smaller than  
 $C_{\rm min}$.
 
 \subsection{Induced traffic breakdown as one of  different consequences of spillover in real traffic \label{Spillover_S}}
  
The effect of continuous upstream propagation of traffic congestion is often 
called   {\it spillback}. When due to this   upstream propagation a
congested pattern affects an upstream  road bottleneck, it is often called {\it spillover}.
  In the cases of the wide moving jams
shown in Figs.~\ref{Breakdown} (c) and~\ref{Breakdown} (d), any of 
the jams can also be considered the  effect of spillover because the jam forces
congested traffic at the bottleneck.

However, when the jams are far away upstream of the bottleneck, they do not force congested traffic
  at the bottleneck any more. We can see that there can be at least the following qualitatively different effects
  due to   spillover at a highway bottleneck:
  
(i)  An empirical induced traffic breakdown occurs due to   jam propagation through a bottleneck (Fig.~\ref{Breakdown} (c)).

(ii) An expanded congested pattern (EP) occurs due to spillover (Fig.~\ref{23061998S_23032001N} (b))~\cite{Kerner2002}: The EP shown in Fig.~\ref{23061998S_23032001N} (b) appears when
 an empirical congested pattern    that occurs initially at an off-ramp bottleneck  
  propagates upstream  (spillback). Due to this upstream pattern propagation
  it forces congested conditions at an upstream on-ramp bottleneck   (labeled by $\lq\lq$on-ramp bottleneck 1"); this
 spillover lasts several hours. This case of spillover cannot be considered as induced traffic breakdown, because
   congested traffic at the on-ramp bottleneck is forced
by spillover.

(iii)   The jam propagation
through a bottleneck leads neither to induced traffic breakdown nor to an EP (Fig.~\ref{Breakdown} (d)).
  This effect of spillover shows that   the flow rate is smaller than the minimum capacity of free flow at the bottleneck: $q< C_{\rm min}$.

 \subsection{Conclusions}
 
 An empirical study of real field  
traffic data   
allows us to make the following conclusions about   physical features of empirical nuclei for spontaneous traffic breakdown in free flow
at  highway bottlenecks:

1. In the most real field traffic data measured   in 1996--2014  by road detectors on German freeways,   a  nucleus for traffic breakdown at a highway bottleneck 
occurs  through an interaction of one of the waves in free flow with a  permanent   speed disturbance localized at 
  a highway bottleneck. When the wave reaches the  location of the disturbance at the bottleneck (effective bottleneck location), spontaneous traffic breakdown, i.e.,
 phase transition from free flow
to synchronized flow    occurs.

2. Waves in free flow, which can be   nuclei for spontaneous traffic breakdown at 
 highway  bottlenecks, appear
  due to  oscillations in the percentage of slow   vehicles over time. 
  These  waves  
  propagate with the average speed of   slow   vehicles in free flow (about 85--88 km/h for German highways).
  Within a wave,   the total flow rate is larger and
 the   speed averaged across the highway is smaller    than outside the wave.

 3.  Any of the waves in free flow, which can be a  nucleus for spontaneous traffic breakdown at 
a highway  bottleneck, exhibits
 a  two-dimensional (2D) asymmetric spatiotemporal  structure   whose characteristics 
 are different in different highway lanes.

 4. Microscopic traffic simulations with a stochastic traffic flow model in the framework of three-phase theory
  explain the empirical findings.

 \appendix
 \section{Stochastic three-phase traffic flow model used for simulations
  \label{D_Model}}

 \subsection{Update  rules of vehicle motion \label{Model_U_R}} 
 
 The   traffic flow model used in Sec.~\ref{Phys_S1} (see Tables~\ref{table_CA} and~\ref{table_CA_lane_change})~\cite{KKl2010A} is a discrete version~\cite{KKl2009A} of  the  
   stochastic three-phase traffic flow model of Ref.~\cite{KKl,KKl2003A}:
   rather than the continuum space co-ordinate,   
a discretized space co-ordinate with a small enough value  of the discretization cell $\delta x$  is used. Consequently,  
  the vehicle speed and acceleration (deceleration) discretization intervals are $\delta v$= $\delta x/\tau$
 and   $\delta a$= $\delta v/\tau$, respectively, where time step $\tau=$ 1 s. 

In formulae of Tables~\ref{table_CA} and~\ref{table_CA_lane_change}, $n=0, 1, 2, ...$ is number of time steps,
  $x_{n}$ is the vehicle coordinate 
at time step $n$, $v_{n}$ is the vehicle speed at time step $n$, 
$v_{\rm free}$ is a maximum speed in free flow,
$g_{n}=x_{\ell, n}-x_{n}-d$ is a space gap,
$d$ is a vehicle length, 
  index  $\ell$  marks the preceding vehicle,
$G(v_{n}, v_{\ell,n})$ is a synchronization gap,
 superscripts $+$  and  $-$  in variables,  parameters, and functions 
denote  the preceding vehicle  and  the trailing vehicle 
in the $\lq\lq$target"  (neighboring) lane,  respectively;  the target lane is the 
lane  into which  the vehicle wants to change.

 Because in 
  the discrete model version     discretized (and dimensionless) speed and acceleration 
 are used, which are measured respectively in the discretization values $\delta v$ and  $\delta a$,
 the value $\tau$ in all formulae below is assumed to be the dimensionless value $\tau=1$.

 \begin{table}
\caption{Discrete version of stochastic three-phase traffic flow  model of Ref.~\cite{KKl2003A}}
\label{table_CA}
\begin{center}
\begin{tabular}{|l|}
\hline
\multicolumn{1}{|c|}{Vehicle motion in road lane} \\
\hline
\multicolumn{1}{|c|}{
$v_{ n+1}=\max(0, \min({v_{\rm free}, \tilde v_{ n+1}+\xi_{ n}, v_{ n}+a
\tau, v_{{\rm s},n} }))$,
}\\
\multicolumn{1}{|c|}{
$x_{n+1}= x_{n}+v_{n+1}\tau$,
}\\
\multicolumn{1}{|c|}{
$\tilde v_{n+1}=\min(v_{\rm free},  v_{{\rm s},n}, v_{{\rm c},n})$,
} \\
\multicolumn{1}{|c|}{
$v_{{\rm c},n}=\left\{\begin{array}{ll}
v_{ n}+\Delta_{ n} &  \textrm{at $g_{n} \leq G_{ n}$,} \\
v_{ n}+a_{ n}\tau &  \textrm{at $g_{n}> G_{ n}$}, \\
\end{array} \right.$
} \\
\multicolumn{1}{|c|}{
$\Delta_{ n}=\max(-b_{ n}\tau, \min(a_{ n}\tau, \ v_{ \ell,n}-v_{ n}))$, 
} \\
\multicolumn{1}{|c|}{
$g_{n}=x_{\ell, n}-x_{n}-d$,
} \\
$v_{\rm free}$, $a$, and $d$   are constants, $\tau=1$; \\
\hline  
\multicolumn{1}{|c|}{Stochastic time delay of acceleration and
deceleration:} \\
\multicolumn{1}{|c|}{$a_{n}=a  \Theta (P_{\rm 0}-r_{\rm 1})$, \
$b_{n}=a  \Theta (P_{\rm 1}-r_{\rm 1})$,} \\
\multicolumn{1}{|c|}{
$P_{\rm 0}=\left\{
\begin{array}{ll}
p_{\rm 0} & \textrm{if $S_{ n} \neq 1$} \\
1 &  \textrm{if $S_{ n}= 1$},
\end{array} \right.
\quad
P_{\rm 1}=\left\{
\begin{array}{ll}
p_{\rm 1} & \textrm{if $S_{ n}\neq -1$} \\
p_{\rm 2} &  \textrm{if $S_{ n}= -1$},
\end{array} \right.$
}\\
\multicolumn{1}{|c|}{
$S_{ n+1}=\left\{
\begin{array}{ll}
-1 &  \textrm{if $\tilde v_{ n+1}< v_{ n}$} \\
1 &  \textrm{if $\tilde v_{ n+1}> v_{ n}$} \\
0 &  \textrm{if $\tilde v_{ n+1}= v_{ n}$},
\end{array} \right.$
}\\
$\Theta (z) =0$ at $z<0$ and $\Theta (z) =1$ at $z\geq 0$, \\
$r_{1}={\rm rand}(0,1)$, $p_{\rm 0}=p_{\rm 0}(v_{n})$,  $p_{\rm 2}=p_{\rm 2}(v_{n})$,
 $p_{\rm 1}$ is constant. \\
\hline  
\multicolumn{1}{|c|}{Model speed fluctuations:} \\
\multicolumn{1}{|c|}{
$\xi_{ n}=\left\{
\begin{array}{ll}
\xi_{\rm a} &  \textrm{if  $S_{ n+1}=1$} \\
- \xi_{\rm b} &  \textrm{if $S_{ n+1}=-1$} \\
\xi^{(0)} &  \textrm{if  $S_{ n+1}=0$},
\end{array} \right.$
}\\
\multicolumn{1}{|c|}{$\xi_{\rm a}=a^{(\rm a)} \tau \Theta (p_{\rm a}-r)$, \
$\xi_{\rm b}=a^{(\rm b)} \tau \Theta (p_{\rm b}-r)$,} \\
\multicolumn{1}{|c|}{
$\xi^{(0)}=a^{(0)}\tau \left\{
\begin{array}{ll}
-1 &  \textrm{if $r\leq p^{(0)}$} \\
1 &  \textrm{if $p^{(0)}< r \leq 2p^{(0)}$ and $v_{n}>0$} \\
0 &  \textrm{otherwise},
\end{array} \right.$
}\\
$r={\rm rand}(0,1)$;
$a^{(\rm a)}=a^{(\rm a)} (v_{n})$, $a^{(\rm b)}=a^{(\rm b)} (v_{n})$; \\
   $p_{\rm a}$, $p_{\rm b}$, $p^{(0)}$, 
 $a^{(0)}$ 
are constants.\\
\hline  
\multicolumn{1}{|c|}{Synchronization gap $G_{n}$ and safe speed $v_{{\rm s},n}$:} \\
\multicolumn{1}{|c|}{
$G_{n}=G(v_{n}, v_{\ell,n})$,
} \\
\multicolumn{1}{|c|}{
$G(u, w)=\max(0,  \lfloor k\tau u+  a^{-1}\phi_{0}u(u-w) \rfloor),$
} \\
\multicolumn{1}{|c|}{
$v_{{\rm s},n}=
\min{(v^{\rm (safe)}_{ n},  g_{ n}/ \tau  + v^{\rm (a)}_{ \ell})}$,  
} \\
\multicolumn{1}{|c|}{
$v^{\rm (safe)}_{ n}=\lfloor v^{\rm (safe)} (g_{n}, \ v_{ \ell,n}) \rfloor$, 
} \\
 \multicolumn{1}{|c|}{
$v^{\rm (safe)} \tau_{\rm safe} + X_{\rm d}(v^{\rm (safe)}) = g_{n}+X_{\rm d}(v_{\ell, n})$,
} \\
 \multicolumn{1}{|c|}{
$X_{\rm d} (u)=b \tau^{2} \bigg(\alpha \beta+\frac{\alpha(\alpha-1)}{2}\bigg)$, $\alpha=\lfloor u/b\tau \rfloor$, $\beta=u/b\tau-\alpha$,
} \\
\multicolumn{1}{|c|}{
$v^{\rm (a)}_{\ell}=
\max(0, \min(v^{\rm (safe)}_{ \ell, n}, v_{ \ell,n},  g_{ \ell, n}/\tau)  -a\tau),$
} \\
 $\tau_{\rm safe}$
 is a safe time gap; 
$b$, $k>1$, and $\phi_{0}$ are constants; \\ $\lfloor z \rfloor$ denotes the  integer part of a real number $z$. \\
\hline
\end{tabular}
\end{center}
\end{table}
\vspace{1cm}

 \begin{table}
\caption{Lane changing rules in discrete version of stochastic three-phase traffic flow  model of Ref.~\cite{KKl2003A}}
\label{table_CA_lane_change}
\begin{center}
\begin{tabular}{|l|}
\hline
\multicolumn{1}{|c|}{Lane changing rules}  \\
\hline
\multicolumn{1}{|c|}{Lane changing occurs with probability $p_{\rm c}$}  \\
\multicolumn{1}{|c|}{from right to left lane $R \rightarrow L$ and  back $L\rightarrow R$:} \\
\hline
\multicolumn{1}{|c|}{Incentive conditions for lane  changing} \\
\multicolumn{1}{|c|}{
$R \rightarrow L$: $v^{+}_{n} \geq v_{\ell, n}+\delta_{1}$   and $v_{n}\geq v_{\ell, n}$,
}\\
\multicolumn{1}{|c|}{
$L \rightarrow R$: $v^{+}_{n} > v_{\ell, n}+\delta_{1}$ or $v^{+}_{n}>v_{n}+\delta_{1}$.
}\\
In  conditions $R \rightarrow L$ and $L \rightarrow R$,   
the value $v^{+}_{n}$ at $g^{+}_{n}>L_{\rm a}$  \\ 
and the value  $v_{\ell, n}$ at $g_{n}>L_{\rm a}$ are replaced by $\infty$,  \\
where $L_{\rm a}$ is constant. \\
\hline
\multicolumn{1}{|c|}{Safety conditions for lane  changing} \\ 
\multicolumn{1}{|c|}{
rules ($\ast $):  $g^{+}_{n} >\min(v_{n}\tau, \ G^{+}_{n})$, \ $g^{-}_{n} >\min(v^{-}_{n}\tau, \ G^{-}_{n})$,  
}\\
\multicolumn{1}{|c|}{
\  where  
$G^{+}_{n}=G( v_{n}, v^{+}_{n})$,
 $G^{-}_{n}=G(v^{-}_{n}, v_{n})$, 
 }\\ 
 \multicolumn{1}{|c|}{{\it or}   } \\
 \multicolumn{1}{|c|}{
rule ($\ast\ast $):  $x^{+}_{n}-x^{-}_{n}-d > g^{\rm (min)}_{\rm target}$ \ with \ $g^{\rm (min)}_{\rm target}=\lfloor \lambda  v^{+}_{n} +d \rfloor$,
 }\\
 the vehicle should pass the midpoint point $x^{\rm (m)}_{n}$ \\
 between   two neighboring vehicles in the target lane. \\
\hline 
 \multicolumn{1}{|c|}{Speed after lane changing} \\
\multicolumn{1}{|c|}{ $v_{ n}=\hat v_{n}$, \ $\hat v_{n}=  \min( v^{ +}_{n},  \ v_{n}+\Delta v^{(1)})$,
}\\ 
in $\hat v_{n}$ the speed $v_{n}$ is  related to    the initial lane  \\ before lane changing. \\
 \multicolumn{1}{|c|}{Vehicle coordinate after lane changing} \\
  does not changes under   rules ($\ast $) \\ and it   changes to $x_{n}=x^{\rm (m)}_{n}$
under   rule ($\ast\ast $).  
 \\
\hline
\end{tabular}
\end{center}
\end{table}
\vspace{1cm}

\subsection{Model of vehicle merging at moving bottleneck   \label{Model_S_M}}

In accordance with~\cite{KKl2010A}, we assume that a slow vehicle moves in the right lane.
If a vehicle moves initially in the right lane upstream   of the slow vehicle, then within the moving merging region
$L_{\rm c}$ (Fig.~\ref{Model} (b)) the vehicle changes from the right lane to the left lane, 
  when    
  safety conditions ($\ast$) {\it or}    ($\ast \ast$)  
  are satisfied.
  
  The safety conditions ($\ast$) are as follows:
  \begin{equation}
  \begin{array}{ll}
g^{+}_{n} >\min(\hat  v_{n}\tau , \ G(\hat  v_{n}, v^{+}_{n})), \\
g^{-}_{n} >\min(v^{-}_{n}\tau, \ G(v^{-}_{n},\hat  v_{n})),
\end{array} 
\end{equation}
\begin{equation}
 \hat v_{n}=\min(v^{+}_{n},  \ v_{n}+\Delta v^{(1)}_{r}), 
 \label{A2}
\end{equation}
 $\Delta v^{(1)}_{r}>0$ is constant
 
 The safety condition  ($\ast \ast$) is given by formula
\begin{equation}
x^{+}_{n}-x^{-}_{n}-d > g^{\rm (min)}_{\rm target}, \quad g^{\rm (min)}_{\rm target}=\lfloor \lambda_{\rm b}  v^{+}_{n} +d \rfloor,
\label{merging_b}
\end{equation}
 $\lambda_{\rm b}$ is constant; in addition,  
the vehicle should pass the midpoint $ x^{\rm (m)}_{n}=\lfloor (x^{+}_{n}+x^{-}_{n})/2 \rfloor$ 
between two neighboring vehicles in the target lane, i.e., the conditions 
  \begin{equation}
 \begin{array}{ll}
x_{n-1}< x^{\rm (m)}_{n-1} \  \textrm{and} \
 x_{n} \geq x^{\rm (m)}_{n} \\
\ \textrm{\it or} \\
x_{n-1} \geq x^{\rm (m)}_{n-1} \  \textrm{and} \
 x_{n} < x^{\rm (m)}_{n}.
\end{array} 
\label{mid2}
\end{equation}
should be satisfied.

  Speed adaptation before vehicle merging is given by vehicle motion rules of Tables~\ref{table_CA},
where
\begin{equation}
 v_{{\rm c},n}=
 \left\{\begin{array}{ll}
v_{ n}+\Delta^{+}_{ n}   \  \textrm{at} \
g^{+}_{n} \leq G(v_{n}, \hat
v^{+}_{n})  \\
v_{ n}+a_{ n}\tau  \  \textrm{at} \ g^{+}_{n}>G( v_{n}, \hat
v^{+}_{n}) 
\end{array}\right.  
\end{equation}
\begin{equation}
 \Delta^{+}_{ n}=\max(-b_{ n}\tau, \min(a_{ n}\tau, \ \hat v^{+}_{n}-v_{
n})), 
\end{equation}
\begin{equation}
 \hat v^{+}_{n}=\max(0, \min(v_{\rm free}, \  v^{+}_{n}+\Delta
v^{(2)}_{r})), 
\end{equation}
$\Delta v^{(2)}_{r}$ is  constant. 

After vehicle merging
the vehicle speed $v_{ n}$ is  set to $\hat v_{ n}$ (\ref{A2}); note that
in (\ref{A2}) 
 the vehicle speed $v_{n}$ is the speed   before vehicle merging;
 under the rule ($\ast $) the vehicle coordinate $x_{n}$  does not change,
under the rule ($\ast \ast$): 
\begin{equation}
x_{n} = x^{\rm
(m)}_{n}.
\end{equation}

 The same rules for vehicle merging are used in models of    an on-ramp
 bottleneck (Fig.~\ref{Model} (b))  (see Fig.~16.2 in Sect.~16.3.6 of~\cite{KernerBook}), i.e.,
 when a vehicle merges from the on-ramp onto the main road or a vehicle leaves the main road
 to the off-ramp.

\begin{table}
\caption{Model parameters used in most  simulations (when other model parameters or the continuum stochastic
model of~\cite{KKl2003A} are used, this is mentioned in
figure captions)}
\label{table2}
\begin{center}
\begin{tabular}{|l|}
\hline
\multicolumn{1}{|c|}{Vehicle motion in road lane:
}\\
\hline
$\tau_{\rm safe}   = \tau$, $d = 7.5 \  \rm m/\delta x$, $\delta x=$ 0.01 m, \\
$v^{\rm (max)}_{\rm free} = 30 \ {\rm ms^{-1}}/\delta v$, $b = 1 \ {\rm ms^{-2}}/\delta a$, \\ $\delta v= 0.01 \  {\rm ms^{-1}}$, \ $\delta a= 0.01 \  {\rm ms^{-2}}$,
$k=$ 3, \\ $p_{1}=$ 0.3, $\phi_{0}=1$,
 $p_{b}=   0.1$, $p_{a}=   0.17$ \\
$p^{(0)}= 0.005$, 
$p_{\rm 2}(v_{n})=0.48+ 0.32\Theta{( v_{n}-v_{21})}$, \\
$p_{\rm 0}(v_{n})=0.575+ 0.125\min{(1, v_{n}/v_{01})}$, \\
 $a^{(\rm b)}(v_{n})=0.2a+$ \\
  $+0.8a\max(0, \min(1, (v_{22}-v_{n})/\Delta v_{22})$, \\
  $a^{(0)}= 0.2a$, 
  $a^{(\rm a)}= a$, \\
  $v_{22} = 12.5 \ {\rm ms^{-1}}/\delta v$,  
  $\Delta v_{22} = 2.778 \ {\rm ms^{-1}}/\delta v$, \\
$v_{01} = 10 \ {\rm ms^{-1}}/\delta v$, $v_{21} = 15 \ {\rm ms^{-1}}/\delta v$, $a=$ 0.5 ${\rm ms^{-2}}/\delta a$. \\
\hline
\multicolumn{1}{|c|}{Lane changing:}
\\
\hline  
$\delta_{1}=1$  
  ${\rm ms^{-1}}/\delta v$,  
   $L_{\rm a}=80 \  {\rm m}/\delta x$, \\
  $p_{\rm c}=0.2 $, 
 $\lambda=0.75$, 
   $\Delta v^{(1)}=2$
   ${\rm ms^{-1}}/\delta v$. \\
   \hline
    \multicolumn{1}{|c|}{Bottleneck models:}
\\
 \hline
$\lambda_{\rm b}=$ 0.75 for all the bottlenecks, \\
$L_{\rm c}= 0.3 \ {\rm km}/\delta x$ 
   for moving bottleneck, \\
   $v_{\rm free \ on}=22.2 \ {\rm ms^{-1}}/\delta v$, \\ 
   $\Delta v^{\rm (2)}_{\rm r}=$ 5    ${\rm ms^{-1}}/\delta v$ for on-ramp bottleneck, \\
   $L_{\rm r}=1 \ {\rm km}/\delta x$,   $\Delta v^{\rm (1)}_{\rm r}=10 \ {\rm ms^{-1}}/\delta v$,  \\
   $L_{\rm m}=$ 0.3   \ ${\rm km}/\delta x$  for on-ramp bottleneck. \\
\hline
\end{tabular}
\end{center}
\end{table}
\vspace{1cm}

 \section{Model of moving virtual detector   \label{Sim_MB_S}}

    To reconstruct a 2D-structure of the flow rate (Fig.~\ref{Model} (d)), we
use a moving virtual detector at which the vehicle speed $v=v(x, \ t)$ 
and   headway between vehicles $h=h(x, \ t)$ are averaged in a neighborhood of the moving bottleneck  as follows:
\begin{equation}
v_{\rm avr} (x, \ t)=\frac{1}{T}\int^{t+T/2}_{t-T/2} {v(x+v_{\rm det}(t^{\prime}-t), \ t^{\prime})dt^{\prime}},
\label{speed_avr}
\end{equation}
\begin{equation}
h_{\rm avr} (x, \ t)=\frac{1}{T}\int^{t+T/2}_{t-T/2} {h(x+v_{\rm det}(t^{\prime}-t), \ t^{\prime})dt^{\prime}},
\label{headway_avr}
\end{equation}
where $T=$ 5 min is the  averaging time interval, 
   $v_{\rm det}$
 is the speed of virtual detector (we use
    $v_{\rm det}=v_{\rm M}$).
The flow rate is 
\begin{equation}
    q_{\rm avr} (x, \ t)=
v_{\rm avr} (x, \ t)/h_{\rm avr} (x, \ t).
\label{flow_avr}
\end{equation}In the discrete form,  coordinate $x$ and time $t$ are discrete ones:
$x_{m}=m h_{\rm x}$ and $t_{k}=k h_{\rm t}$, $m=0,1,2,...$, $k=0,1,2,...$,
where the space step $h_{\rm x}=$7.5 m  and the time step $h_{\rm t}=$1 s.
Then formulae (\ref{speed_avr}), (\ref{headway_avr}) take the form
\begin{equation}
v_{\rm avr} (x_m, \ t_k)=\frac{1}{N_T+1}\sum^{k^{\prime}=N_{\rm T}/2}_{k^{\prime}=-N_{\rm T}/2}
 {v(x_m+h_{\rm x}m^{\prime}, \ t_k+ h_{\rm t}k^{\prime})},
\label{speed_avr_disc}
\end{equation}
\begin{equation}
h_{\rm avr} (x_m, \ t_k)=\frac{1}{N_T+1}\sum^{k^{\prime}=N_{\rm T}/2}_{k^{\prime}=-N_{\rm T}/2}
 {h(x_m+h_{\rm x}m^{\prime}, \ t_k+ h_{\rm t}k^{\prime})},
\label{headway_avr_disc}
\end{equation}
where
\begin{equation}
m^{\prime}=\lfloor v_{\rm det} h_{\rm t}k^{\prime}/h_{\rm x}\rfloor,
\label{number_x}
\end{equation}
the expression $\lfloor z \rfloor$ denotes the integer part of $z$, 
the value $N_{\rm T}$  is chosen to be the even value and $T=(N_{\rm T}+1) h_{\rm t}$.


\begin{thebibliography}{8.}
\bibitem{FlowIncrease}

The effective bottleneck location is a road location at which
 traffic breakdown is observed ($x\approx $ 6.4 km).
In the reality, the on-ramp bottleneck shown in Fig.~\ref{Breakdown} (a)  consists
of two different on-ramps
that are in intersection $\lq\lq$Friedberg" of freeway A5-South with other
road(s). Explanations of this effective on-ramp bottleneck can be found 
in Sec.~2.3.2 and 9.2.1 of the book~\cite{KernerBook}. 
The inflow   to the upstream on-ramp explains   
  the increase in the flow rates $q$ and $q_{\rm slow}$ 
 occurring  about 0.5--1 km upstream of the effective bottleneck location.
\bibitem{Permanent}
  It can be assumed that a permanent local speed
disturbance  in free flow at the effective location of an on-ramp bottleneck is associated with vehicles merging from on-ramp lane to the main 
road~\cite{KernerBook,KernerBook2}.
These vehicles force the vehicles on the main road to decrease their speed in a neighborhood of the on-ramp merging region.
This assumption is confirmed by simulations made in Sec.~\ref{Phys_D_S2} (see Fig.~\ref{Theory3}).
It can also be assumed that a permanent local speed
disturbance  in free flow at the effective location of an off-ramp bottleneck is associated with 
vehicles that must change the lane to leave the main road to the off-ramp.
This lane changing, which occurs mostly in a small road region downstream of the off-ramp lane, causes the speed decrease
of vehicles on the main road. 
However, to prove these theoretical assumptions with empirical data,   measurements of
microscopic (single-vehicle) spatiotemporal
data (e.g., vehicle trajectories)
of almost all vehicles moving in free flow in a neighborhood of a highway bottleneck are required.
Unfortunately, such   data is not currently available.
\end{thebibliography}
\end{document}